\documentclass[aps,prd,reprint,twocolumn,preprintnumbers,floatfix,nofootinbib]{revtex4-1}
\usepackage{booktabs}
\usepackage[english]{babel}
\usepackage{amsmath,amssymb,amsbsy,amstext, amsthm, simplewick}
\usepackage{hyperref}
\usepackage{graphicx}
\usepackage{amsfonts}
\usepackage{amssymb}
\usepackage[small]{caption}
\usepackage{upgreek}
\usepackage{framed}
\usepackage{enumitem}
\usepackage{soul}

\usepackage{booktabs}
\usepackage[english]{babel}
\usepackage{amsmath,amssymb,amsbsy,amstext, amsthm, simplewick}
\usepackage{wrapfig}
\usepackage{hyperref}
\usepackage{graphicx}
\usepackage{amsfonts}
\usepackage{amssymb}
\usepackage{subfig}
\usepackage{upgreek}
 \usepackage{exscale,relsize}
 \usepackage[makeroom]{cancel}
\usepackage{soul}
\usepackage{bbold}
\usepackage{lipsum}
\usepackage{mdframed}
\usepackage{mathtools}
\usepackage[dvipsnames]{xcolor}
\usepackage{tikz}
\usepackage{tikz-cd}
\usepackage[export]{adjustbox}
\usepackage{dsfont}
\usepackage[mathscr]{eucal}

 \newcommand{\circled}[2][]{%
  \tikz[baseline=(char.base)]{%
    \node[shape = circle, draw, inner sep = 1pt]
    (char) {\phantom{\ifblank{#1}{#2}{#1}}};%
    \node at (char.center) {\makebox[0pt][c]{#2}};}}
\robustify{\circled}

\newcommand*\diff{\mathop{}\!\mathrm{d}}

\usepackage{colortbl}
\definecolor{lightgreen}{cmyk}{0.2, 0, 0.2, 0.2}
\definecolor{lightgray}{cmyk}{0.1,0.2,0,0.1}
\definecolor{lightgray2}{cmyk}{0.1,0.1,0,0.1}

\makeatletter
\newlength{\apb@width}
\newcommand{\autoparbox}[2][c]{\settowidth{\apb@width}{#2}\parbox[#1]{\apb@width}{#2}}

\newcommand{\Cen}[2]{%
  \ifmeasuring@
    #2%
  \else
    \makebox[\ifcase\expandafter #1\maxcolumn@widths\fi]{$\displaystyle#2$}%
  \fi
}
\makeatother

\newcommand{\beq}{\begin{equation}\begin{aligned}}
\newcommand{\eeq}{\end{aligned}\end{equation}}

\hypersetup{
    colorlinks,
    linkcolor={red!50!black},
    citecolor={blue!50!black},
    urlcolor={blue!80!black}
}

\def\beq{\begin{equation}}
\def\eeq{\end{equation}}

\def\Beq{\begin{equation}\begin{aligned}}
\def\Eeq{\end{aligned}\end{equation}}

\def\bea{\begin{eqnarray}}
\def\eea{\end{eqnarray}}

\def\beq{\begin{equation}}
\def\eeq{\end{equation}}
\def\bea{\begin{eqnarray}}
\def\eea{\end{eqnarray}}

\def\bp{\boldsymbol{p}}

\def\bk{\boldsymbol{k}}

\def\bx{\boldsymbol{x}}
\def\bq{\boldsymbol{q}}

\DeclareRobustCommand{\SkipTocEntry}[4]{}
\DeclareSymbolFont{extraup}{U}{zavm}{m}{n}
\DeclareMathSymbol{\varheart}{\mathalpha}{extraup}{86}
\DeclareMathSymbol{\vardiamond}{\mathalpha}{extraup}{87}
\hypersetup{
    colorlinks,
    linkcolor={red!100!black},
    citecolor={blue!100!black},
    urlcolor={blue!80!black}
}

\begin{document}

\preprint{DESY-23-027}

\title{Isocurvature Constraints on Scalar Dark Matter Production from the Inflaton}

\author{Marcos A.~G.~Garcia$^{a}$}
\email{marcos.garcia@fisica.unam.mx}
\author{Mathias Pierre$^{b}$}
\email{mathias.pierre@desy.de}
\author{Sarunas Verner$^{c}$}
\email{verner.s@ufl.edu}
\vspace{0.5cm}

 \affiliation{
$^a$
Departamento de F\'isica Te\'orica, Instituto de F\'isica, Universidad Nacional Aut\'onoma de M\'exico, Ciudad de M\'exico C.P. 04510, Mexico} 
\affiliation{$^b$ Deutsches Elektronen-Synchrotron DESY, Notkestr. 85, 22607 Hamburg, Germany} 
\affiliation{${}^c$ Institute for Fundamental Theory, Physics Department, University of Florida, Gainesville, FL 32611, USA}

\date{\today}

\begin{abstract} 
We investigate the production of a spectator scalar dark matter field that is directly coupled to the inflaton during inflation and reheating. We consider two specific inflationary potentials, namely the Starobinsky and T-model of inflation, which satisfy the constraints on the scalar tilt, $n_s$, and tensor-to-scalar ratio, $r$, measured by the \textit{Planck} satellite.  Excitation of light scalar dark matter during inflation may result in large isocurvature perturbations, which can be avoided by inducing a sizable effective dark matter mass during the inflationary phase. For purely gravitational production, the \textit{Planck} isocurvature constraints require the dark matter mass to be larger than the Hubble scale at horizon exit, with $m_{\chi} \gtrsim 0.5H_*$. For small bare dark matter masses $m_{\chi} \ll H_*$, these constraints translate into a lower bound on the dark matter coupling to the inflaton. We argue that these constraints can be applied to a wide class of single-field slow-roll inflation models. We also derive isocurvature, dark matter abundance, and Lyman-$\alpha$ constraints on the direct coupling and bare dark matter mass. 
\end{abstract}

\maketitle

%%%%%%%%%%%%%%%%%%%%%%%%%%%%%%%%%%%%%%%%%%%%%%%%
\section{Introduction}
\label{sec:introduction}
%%%%%%%%%%%%%%%%%%%%%%%%%%%%%%%%%%%%%%%%%%%%%%%%
The nature and origin of dark matter (DM) remains one of the biggest puzzles in fundamental physics. The stringent limits from direct DM detection searches, such as XENON1T~\cite{XENON:2018voc}, LUX~\cite{LUX:2016ggv}, PandaX~\cite{PandaX-4T:2021bab} and LZ~\cite{LZ:2022ufs}, along with the absence of detection from indirect DM and collider experiments, create tension with the standard weakly interacting massive particle (WIMP) paradigm without providing clues about the composition of the invisible fraction of the universe. This tension provides a strong incentive to explore alternative models of dark matter~\cite{Arcadi:2017kky,Escudero:2016gzx,Mambrini}.

Among the various proposed dark matter production mechanisms, a particularly minimalistic model assumes that the dark sector of the universe is po\-pu\-la\-ted independently of the visible sector, but via the same mechanism. That is, the energy density stored in the degrees of freedom that drive primordial inflation dissipates during the post-inflationary reheating epoch and populates the dark sector via the \textit{freeze-in} mechanism~\cite{Hall:2009bx, Bernal:2017kxu}. If the dark sector only interacts directly with the inflaton field, dark matter would never have been in thermal equilibrium with Standard Model particles, and this could lead to non-perturbative gravitational particle production~\cite{Parker:1968mv, Parker:1969au, Parker:1971pt, Ford:1986sy, Ema1, Ema2, Ema3, Herranen:2015ima, Markkanen:2015xuw}. Therefore, its comoving phase space distribution (PSD) would be fully determined by the end of reheating.

Our previous study~\cite{Garcia:2022vwm} (see also~\cite{Chung:1998zb, Chung:1998ua, Kuzmin:1998kk, Ling:2021zlj, Kaneta:2022gug}) explored this minimal out-of-equilibrium scenario for a scalar DM particle. We determined the relic abundance and structure formation constraints for a wide range of inflaton-DM coupling values, covering pure gravitational production, weak direct coupling (perturbative production), and strong direct coupling (non-perturbative production) regimes. In the weak coupling regime, we found that Lyman-$\alpha$ forest measurements of the matter power spectrum allow DM masses as small as $\sim 0.3 \, {\rm meV}$. However, purely gravitationally produced dark matter faces a tension with the Cosmic Microwave Background (CMB) bounds on the amplitude of the isocurvature power spectrum for DM masses smaller than the instantaneous Hubble scale at the end of inflation~\cite{Chung:2004nh, Chung:2011xd, Chung:2015pga, Herring:2019hbe, Padilla:2019fju, Ling:2021zlj, Redi:2022zkt}. The current constraints on the isocurvature power spectrum from ${\textit Planck}$ are given by $\beta_{\rm iso}\;\equiv\; \mathcal{P}_{\mathcal{S}}(k_*)/( \mathcal{P}_{\mathcal{R}}(k_*) + \mathcal{P}_{\mathcal{S}}(k_*) )\;<\;0.038$ at the 95\% C.L. with the pivot scale $k_*=0.05\,{\rm Mpc}^{-1}$~\cite{Planck:2018jri}, where $ \mathcal{P}_{\mathcal{R}}$ and $ \mathcal{P}_{\mathcal{S}}$ denote the curvature and isocurvature power spectra, respectively.

As argued in~\cite{Gordon:2000hv, Herring:2019hbe}, isocurvature modes are always strongly suppressed relative to adiabatic modes at linear order when there is no dark matter misalignment. A cursory computation shows that the curvature perturbation $\mathcal{R}$ is conserved in superhorizon scales, consistent with the absence of late-time isocurvature~\cite{Renaux-Petel:2014htw}. However, during inflation, the tachyonic instability of light DM modes can efficiently source isocurvature modes at second order~\cite{Chung:2004nh, Chung:2011xd, Ling:2021zlj}. In our previous work~\cite{Garcia:2022vwm}, we postponed the detailed exploration of power spectrum constraints for the direct decay production mechanism. In this paper, we investigate the constraints on the dark matter abundance, the Lyman-$\alpha$ forest, and the isocurvature at quadratic order for this out-of-equilibrium DM production scenario.

This paper is structured as follows: Section~\ref{sec:dmgravprod} discusses the non-perturbative gravitational production of scalar dark matter during inflation and reheating. In Section~\ref{sec:dmpsd}, we introduce the phase space distribution of DM and compute the dark matter relic abundance. Section~\ref{sec:isocurvature} is dedicated to computing the dark matter isocurvature power spectrum, and in Section~\ref{sec:relconstraints}, we combine the relic abundance, structure formation, and isocurvature constraints. Finally, Section~\ref{sec:conclusions} summarizes our findings. In this paper, we use natural units $k_B = \hbar = c = 1$ and the metric signature $(+,-,-,-)$. 
%%%%%%%%%%%%%%%%%%%%%%%%%%%%%%%%%%%%%%%%%%%%%%%%%%%%%%%%%%%%%%%%%%%%%%%%%
\section{Inflation and Dark Matter Production}
\label{sec:dmgravprod}
%%%%%%%%%%%%%%%%%%%%%%%%%%%%%%%%%%%%%%%%%%%%%%%%%%%%%%%%%%%%%%%%%%%%%%%%%
We examine the production of a scalar spectator field, $\chi$, that is associated with dark matter. We assume that the scalar inflaton field, $\phi$, can couple to this singlet scalar dark matter particle. For both fields, a minimal coupling to gravity is assumed\footnote{The presence of large non-minimal couplings to gravity in the inflaton or dark matter sector would significantly alter the model, and we leave this analysis for future work.} and a dimension-4 interaction $\frac{1}{2} \sigma \phi^2 \chi^2$. The scalar field action can be expressed as
\begin{align}
\notag
\mathcal{S} \;=\; \int \diff^4x\,&\sqrt{-g} \Big[ \frac{1}{2}(\partial_{\mu} \phi)^2 + \frac{1}{2}(\partial_{\mu}\chi)^2 - V(\phi)\\ \label{eq:action}
&-\frac{1}{2}\left(m_{\chi}^2 + \sigma\phi^2\right)\chi^2 + \mathcal{L}_{\phi-{\rm SM}}\Big]\,.
\end{align}
Here $m_{\chi}$ represents the bare mass of dark matter, $\sigma$ is the dimensionless inflaton-DM coupling, $\mathcal{L}_{\phi-{\rm SM}}$ is the not yet specified inflaton-Standard Model interaction, that is responsible for reheating of the universe, and $V(\phi)$ is the inflaton potential. We consider two particular models of inflation: the Starobinsky model~\cite{Starobinsky:1980te, Ellis:2013nxa} 
\beq
\label{inf:starobinsky}
V(\phi) \;=\; \frac{3}{2}\lambda M_P^4 \left[ 1 - e^{-\sqrt{\frac{2}{3}}\frac{\phi}{M_P}}\right]^2\,,
\eeq
and the T-model~\cite{Kallosh:2013maa}
\beq
\label{inf:tmodel}
V(\phi) \;=\; \lambda M_P^4 \left[ \sqrt{6}\tanh\left(\frac{\phi}{\sqrt{6}M_P}\right)\right]^2 \,,
\eeq
where $M_P=1/\sqrt{8\pi G_N}\simeq 2.435\times 10^{18}\,{\rm GeV}$ denotes the reduced Planck mass and the constant $\lambda$ determines the inflaton mass at its minimum at $\phi=0$, with $m_{\phi} = \sqrt{2\lambda}M_P$ for both models. The value of $\lambda$ is fixed by the measurement of the amplitude of the primordial curvature power spectrum. The principal CMB observables are evaluated at the pivot scale $k_* = 0.05\,{\rm Mpc}^{-1}$ used in the \textit{Planck} analysis~\cite{Planck:2018jri}, and they can be expressed in terms of the slow-roll parameters:
\begin{align}
    n_s &\simeq 1 - 6 \epsilon_* + 2 \eta_* \, , \qquad r \simeq 16 \epsilon_* \, , \\
    A_{S*} &\simeq \frac{V_*}{24\pi^2 \epsilon_* M_P^4} \, .
\end{align}
Here $V_* = V(\phi_*)$, $n_s$ is the scalar tilt, $r$ is the tensor-to-scalar ratio, 
$A_{S*}\simeq 2.1\times 10^{-9}$ is the curvature power spectrum amplitude, and the slow-roll parameters are $\epsilon = \frac{1}{2} M_P^2 (V_{,\phi}/V)^2$ and $\eta = M_P^2 \left(V_{,\phi \phi}/V \right)$. The normalization of the potentials~(\ref{inf:starobinsky}) and~(\ref{inf:tmodel}) can be approximated as~\cite{Garcia:2020wiy, Ellis:2021kad}
\begin{align}
&\lambda \simeq \frac{12\pi^2 A_{S*}}{N_*^2} \;\simeq\; \frac{2.5\times10^{-7}}{N_*^2}~(\rm{Starobinsky})\, , \\
&\lambda \simeq \frac{3\pi^2 A_{S*}}{N_*^2} \;\simeq\; \frac{6.2\times10^{-8}}{N_*^2}~(\rm{T\mbox{-}Model}) \, ,
\end{align}
where $N_*$ is the number of $e$-folds after horizon crossing.\footnote{Note that the dimensionless parameter $\lambda$ uniquely determines the full inflaton potential. For this choice of parameter, the Starobinsky and T-model potentials coincide at large field values. However, the two potentials differ at the end of inflation and around the minimum, resulting in different inflaton masses.} For a nominal choice of $N_* = 55$ $e$-folds, we find $n_s \simeq 0.965~(0.963)$ and $r \simeq 0.004~(0.004)$ for Starobinsky model (T-model) of inflation, and both models are highly favored by current CMB measurements~\cite{Planck:2018jri}.\footnote{See Ref.~\cite{Ellis:2021kad} for a dedicated analysis on the BICEP/\textit{Keck}/WMAP/\textit{Planck} data constraints on the Starobinsky and T-models of inflation.}

During inflation, the coupling of the dark matter scalar to gravity and the inflaton results in the non-adiabatic variation of the dark matter effective mass, leading to subsequent particle production~\cite{Parker:1968mv, Parker:1969au, Parker:1971pt, Ford:1986sy}. After the end of inflation, the scalar inflaton field, $\phi$, undergoes coherent oscillations about the minimum of its potential, and the effective mass of dark matter oscillates accordingly. This leads to additional post-inflationary production of dark matter.\footnote{See Refs.~\cite{Opferkuch:2019zbd, Mambrini:2021zpp, Haque:2021mab, Clery:2021bwz, Clery:2022wib, Barman:2021ugy, Barman:2022tzk, Haque:2022kez} for the treatment of gravitational production of dark matter during reheating.} As the expansion rate of the universe rapidly slows down at this time, the inflaton decays into elementary particles, reheating the universe.

Assuming that there is no initial misalignment for $\chi$, the population of the dark matter relic abundance will be sourced by the growth in the $\chi$ fluctuation modes. For convenience, we use conformal time $\tau$, which is related to cosmic time $t$ as $\diff t/\diff \tau=a$, where $a$ is the scale factor of the flat Friedmann-Robertson-Walker spacetime (FRW) metric. We introduce the rescaled dark matter field
\beq
X \;\equiv\; a\chi \,,
\label{eq:rescaledmodefunction}
\eeq
and variation of the action~(\ref{eq:action}) with respect to the DM field $\chi$ leads to the following equation of motion 
\beq\label{eq:eomconf}
\left[ \frac{\diff^2}{\diff\tau^2} - \nabla^2  - \frac{a''}{a} + a^2m_{\chi}^2 + \sigma a^2\phi^2 \right] X(\tau,\bx) \;=\; 0 \, , 
\eeq
where prime denotes the derivative with respect to the conformal time, $\tau$. The momentum modes of the field $X$ are determined from its canonically quantized form
\beq\label{eq:modedef}
X(\tau,\bx) = \int \frac{\diff^3\bk}{(2\pi)^{3/2}}\,e^{-i\bk\cdot\bx} \left[ X_k(\tau)\hat{a}_{\bk} + X_k^*(\tau)\hat{a}^{\dagger}_{-\bk} \right]\,, 
\eeq
where $\bk$ corresponds to the comoving momentum, $k = |\bk|$, and $\hat{a}_{\bk}$ and $\hat{a}^{\dagger}_{\bk}$ are the annihilation and creation operators, respectively, satisfying the canonical commutation relations $[\hat{a}_{\bk},\hat{a}^{\dagger}_{\bk'} ] = \delta(  \bk-\bk')$, $[\hat{a}_{\bk},\hat{a}_{\bk'} ] = [\hat{a}^{\dagger}_{\bk},\hat{a}^{\dagger}_{\bk'} ] = 0$. The canonical commutation relations between the field, $X_k$, and its momentum conjugate, $X_k'$,
are fulfilled by the Wronskian constraint $X_kX^{*\prime}_k - X_k^*X_k' \;=\; i$.
Substituting Eq.~(\ref{eq:modedef}) into Eq.~(\ref{eq:eomconf}), we obtain the equation of motion for the mode functions, $X_k$,
\beq
\label{eq:eomp}
X_k'' + \omega_k^2 X_k \; = \; 0 \,,
\eeq
with the mode frequency given by
\begin{align}
\label{eq:omega2}
\omega_k^2 \;&\equiv\; k^2  - \frac{a''}{a} + a^2m_{\chi}^2 + \sigma a^2\phi^2~\nonumber\\ 
&=\; k^2 + a^2 \left( \frac{R}{6}  + m_{\chi}^2 + \sigma\phi^2 \right) \,,
\end{align}
where $R$ denotes the Ricci scalar. The initial conditions for $X_k$ are chosen to be the positive-frequency Bunch-Davies vacuum
\beq
X_k(\tau_0) \;=\; \frac{1}{\sqrt{2\omega_k}} \,, \ X_k'(\tau_0) \;=\; -\frac{i\omega_k}{\sqrt{2\omega_k}}\,,
\eeq
which are valid for modes that are deep inside the Hubble horizon, with $|\tau_0\omega_k|\gg 1$. The efficiency of DM particle production is determined by the ratio of couplings $\sigma/\lambda$~\cite{Garcia:2021iag, Garcia:2022vwm}. Eq.~(\ref{eq:omega2}) can be written equivalently as
\begin{align}
\label{eq:omegaH}
\omega_k^2 & = k^2 + a^2 m_{\rm{eff}}^2 \, , \\
{\rm{with}}~~m_{\rm{eff}}^2 &\equiv m_{\chi}^2 + \sigma\phi^2 - \left(\frac{1-3w}{2} \right)H^2 \, .
\end{align}
Here $H=\dot{a}/a$ is the Hubble parameter and $w \equiv P/\rho$ is the equation of state parameter, where $P$ and $\rho$ are the background pressure and energy density, respectively. For low momentum (IR) modes, the mode frequency squared may become negative during inflation if $k^2 < a^2 |m_{\rm{eff}}^2|$, leading to a fast mode growth during inflation due to the tachyonic instability. If the DM bare mass is negligible, $m_{\chi}\ll H$, this occurs when
\beq
k^2 \lesssim a^2\left(  \frac{1-3w}{2}H^2 - \sigma\phi^2  \right) \, ,
\eeq
which implies the tachyonic growth for the low momentum (IR) modes when
\beq
\label{eq:sigmainfvev}
\frac{\sigma}{\lambda} \;\lesssim\; \frac{\alpha(1-3w)}{2}\left(\frac{M_P}{\phi}\right)^2~{\rm{with}}~~\alpha \equiv \frac{H^2}{\lambda M_P^2} \, .
\eeq
During inflation, when $t \ll t_{\rm{end}}$, $\alpha \simeq 0.5~(2)$ for the Starobinsky model (T-model); at the end of inflation, when $t = t_{\rm{end}}$,
$\alpha \simeq 0.12~(0.33)$, which implies the tachyonic growth when $\sigma/\lambda \lesssim 0.5~(0.3)$ for the Starobinsky model (T-model). Therefore, for a smaller effective coupling, the DM phase space distribution is highly sensitive to the inflationary dynamics.

If the dimensionless inflaton-DM coupling is zero, $\sigma = 0$, then the scalar dark matter is produced solely through gravitational interactions. We find from Eq.~(\ref{eq:eomp}) that tachyonic mode growth occurs when
\begin{equation}
    k^2 \lesssim a^2 \left(  \frac{1-3w}{2}H^2 - m_{\chi}^2 \right) \, .
\end{equation}
During inflation, $w \simeq -1$, and for low momentum (IR) modes, the tachyonic mode enhancement arises when $m_{\chi}^2 \lesssim 2H^2$.

It is important to note that the growth of scalar fluctuations for scales that exit the horizon 50-60 $e$-folds before the end of inflation can efficiently generate isocurvature perturbations that would leave an imprint in the CMB. In Sections~\ref{sec:isocurvature} and~\ref{sec:relconstraints}, we provide a detailed discussion and calculation of dark matter isocurvature and constrain the associated parameter space. Our results can be readily applied to various plateau-like single-field inflation models, assuming that both dark matter and inflaton fields couple minimally to gravity.

\begin{figure*}[!t]
\centering
    \includegraphics[width=\textwidth]{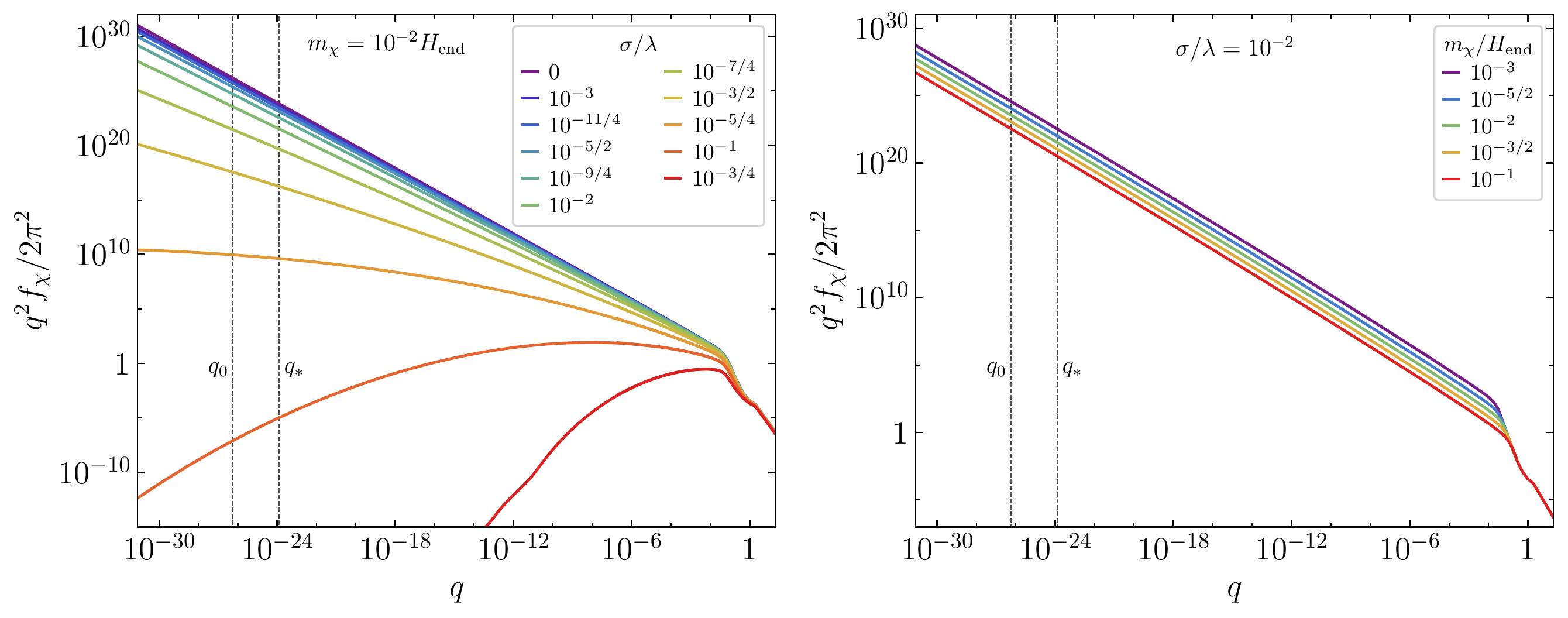}
    \caption{Phase space distribution of the dark matter field $\chi$ for a range of effective couplings $\sigma/\lambda$ and bare masses $m_{\chi}$ for the T-model of inflation. Left panel: the     PSD dependence on the inflaton-DM coupling for a fixed DM mass $m_{\chi} = 10^{-2}H_{\rm{end}}$. As the coupling ratio $\sigma/\lambda$ increases, the long wavelength (IR) modes become more suppressed, and dominant particle production occurs towards the end of inflation. Right panel: the PSD dependence on the DM mass for a fixed inflaton-DM coupling $\sigma/\lambda=10^{-2}$. In the IR regime, $f_{\chi} \propto q^{-2.9}$, and in the UV regime, $f_{\chi} \propto q^{-9/2}$. In both panels, vertical lines are displayed corresponding to the rescaled comoving momenta of the present-day horizon scale, $q_0\simeq 5.8\times 10^{-27}$, and the {\em Planck} pivot scale, $q_*\simeq  1.3\times 10^{-24}$, assuming $N_*=55$. For reference, the horizon scale at the end of inflation is $q_{\rm end}\simeq 0.4$.}
    \label{fig:PSDs}
\end{figure*}
%

%%%%%%%%%%%%%%%%%%%%%%%%%%%%%%%%%%%%%%%%%%%%%%%%%%%%%%%%%%%%
\section{Dark matter phase space distribution}
\label{sec:dmpsd}
%%%%%%%%%%%%%%%%%%%%%%%%%%%%%%%%%%%%%%%%%%%%%%%%%%%%%%%%%%%%
To extract the DM phase space distribution, $f_{\chi}(k,t)$, we numerically solve the equation of motion (\ref{eq:eomp}) for a wide range of momentum modes. This distribution is determined by the comoving particle occupation number of the scalar field, $n_k$, which can be expressed in a UV-finite form~\cite{Kofman:1997yn,Garcia:2021iag}\footnote{This expression can also be equivalently expressed as $n_k = \frac{\omega_k}{2} (|X_k'|^2/\omega_k^2 + |X_k|^2) - \frac{1}{2}$~\cite{Kofman:1997yn}.}
\beq
f_{\chi}(k, t)\;=\; n_{k} \;=\; \frac{1}{2\omega_k}\left| \omega_k X_k - i X'_k \right|^2\,.
\label{eq:PSDandoccupationnumber}
\eeq
This quantity can only be interpreted as the occupation number for non-tachyonic modes, and therefore it must be evaluated well after the end of inflation. For simplicity, we introduce a dimensionless comoving momentum rescaled at the end of inflation with respect to the inflaton mass, $\bq=\bk/(m_{\phi} a_{\rm{end}})$. The comoving number density of the produced DM particles can then be expressed as~\cite{Garcia:2022vwm}
\beq
\label{eq:psdfull}
n_{\chi}\left(\frac{a}{a_{\rm end}}\right)^3 \;=\; \frac{m_{\phi}^3}{2\pi^2}\int \diff q\,q^2 f_{\chi}(q,t) \, .
\eeq
The inflaton mass is given by $m_{\phi} = \sqrt{2 \lambda} M_P \simeq 4.1 H_{\rm{end}} \simeq 3 \times 10^{13} \, \rm{GeV}$ for the Starobinsky model and $m_{\phi} = \sqrt{2 \lambda} M_P = 2.5 H_{\rm{end}}  \simeq 1.6 \times 10^{13} \, \rm{GeV}$ for the T-model.

For concreteness, we assume that the {\em Planck} pivot scale $k_*$ leaves the horizon $N_*=55$ $e$-folds before the end of inflation and a total duration of inflation of $N_{\rm tot}= 76.5$ $e$-folds. We have chosen this value to ensure that the present comoving scale was deeply inside the Hubble horizon at the beginning of inflation, without extending the period of accelerated expansion too far back in time. While the duration of inflation does not affect the form of the PSD due to the linearity of the equation of motion for the momentum modes~(\ref{eq:eomp}), integrated quantities, such as $n_{\chi}$, generally depend on the infrared (IR) cutoff in the spectrum, which we discuss in detail below.

We employ numerical techniques developed in Ref.~\cite{Ellis:2014opa} to compute the power spectra in multifield inflation, to obtain the complete PSD down to wavenumbers below the CMB pivot scale without the need for extrapolation. This represents an improvement over our previous analysis in~\cite{Garcia:2022vwm}. The integrand of Eq.~(\ref{eq:psdfull}) is shown in Fig.~\ref{fig:PSDs} for a range of DM masses and couplings. Although we present the results for the T-model of inflation, they can be translated to other inflationary models by identifying different values of Hubble parameter at the end of inflation between the models.\footnote{This translation is less relevant for the isocurvature constraints since they primarily depend on $H_*$. We discuss it in detail in Section~\ref{sec:isocurvature}.} To translate the constraints from the T-model to the Starobinsky model, we can use the relationship
\beq\label{eq:sigmaH}
\left[\sigma\left(\frac{M_P}{H_{\rm end}}\right)^2 \right]_{\rm T\mbox{-}model} \;=\; \left[\sigma\left(\frac{M_P}{H_{\rm end}}\right)^2\right]_{\rm Staro.}\,,
\eeq
which, in terms of the effective dimensionless coupling ratio, corresponds to $(\sigma/\lambda)_{\rm T\mbox{-}model}\simeq 2.8 (\sigma/\lambda)_{\rm Staro.}$. For the purely gravitational regime, when $\sigma = 0$, the constraints can be directly related by transforming between different values of $H_{\rm{end}}$.\footnote{Note that, for our choice of normalization for the Starobinsky and T-model potentials, the inflaton mass is different in both cases, and the comoving momentum $q$ has to be rescaled accordingly.}

The left panel of Fig.~\ref{fig:PSDs} displays the phase space distributions for a fixed mass $m_{\chi}=10^{-2}H_{\rm end}$ for several values of the effective coupling constant ratio $\sigma/\lambda\ll 1$. For the smallest couplings, the spectrum presents a steep red tilt, corresponding to $f_{\chi}\propto q^{-3}$ in the pure gravitational case ($\sigma=0$). This is consistent with the $k$-dependence of the squared mode function from Eq.~(\ref{eq:modereal}), as derived in Appendix~\ref{app:B}, which is devoted to analytical approximations. For a comprehensive treatment of purely gravitational dark matter production, see also Ref.~\cite{Garcia:2022vwm, Kaneta:2022gug, Ford:2021syk}.

The left panel of Fig.~\ref{fig:PSDs} also illustrates the suppression of the growth of long wavelength (IR) modes for $\sigma\neq0$. This suppression is more significant for modes that exit the horizon earlier because the inflaton field value during inflation is larger, and the tachyonic enhancement is weaker (see Eq.~(\ref{eq:sigmainfvev})). This observation is qualitatively consistent with the analytical approximations provided in App.~\ref{app:B}. Couplings $\sigma/\lambda \gtrsim 10^{-1}$ are not IR-sensitive, and result in particle production that primarily occurs towards the end of inflation and during reheating. As discussed in App.~\ref{app:B}, the phase space distribution in the long wavelength (IR) regime scales as $f_{\chi} \propto q^{-2 \nu}$, where $\nu = \sqrt{9/4 - m_{\rm{eff}}^2/H^2}$ for real $\nu$, and reduces to $f_{\chi} \propto q^{-3}$ when $m_{\rm{eff}} \ll H$, which corresponds to $\sigma \langle \phi \rangle^2 \ll m_{\chi}^2$.

For a non-vanishing coupling $\sigma$, the number density $n_{\chi}$ converges over the full range of momenta $0\leq q<\infty$. However, if we assume a finite duration of inflation, the integrated PSD is sensitive to the IR endpoint for $\sigma/\lambda \lesssim 0.1$, which may be associated with the mode that exited the horizon at the beginning of inflation, or more conservatively, with the wavenumber of the present horizon scale. We choose to ignore the contribution of larger scales that have never entered our causal radius and simply renormalize the background~\cite{Ling:2021zlj,Herring:2019hbe,Starobinsky:1994bd,Felder:1999wt}. 

In all cases, the $q>1$ UV tail of the PSD corresponds to modes excited during the post-inflationary reheating epoch. In the small coupling regime, one can use the perturbative Boltzmann formalism to study how the energy of the inflaton dissipates while it undergoes coherent oscillations around its minimum. The UV tail of the distribution is given by~\cite{Garcia:2022vwm, Kaneta:2022gug}
\begin{align}
\label{eq:psduvtail}
\notag
f_{\chi} (q,t) & \simeq \frac{\sqrt{3}\pi (\sigma - \lambda)^2 \rho_{\phi}^{3/2}(t_{\rm end}) M_P}{16 m_{\phi}^7}\\
&\times q^{-9/2} e^{-1.56\left(\frac{a_{\rm end}}{a_{\rm reh}}\right)^2q^2}~~~(q>1) \, ,
\end{align}
where $\rho_{\phi}(t_{\rm{end}})$ is the inflaton energy density at the end of inflation and $a_{\rm{reh}}$ is the scale factor when reheating occurs.\footnote{We define the time of reheating as the moment when the energy density of radiation starts to dominate the energy budget of the universe.}

The right panel of Fig.~\ref{fig:PSDs} demonstrates the impact of varying the bare mass of the DM field for a fixed inflaton-DM coupling of $\sigma/\lambda = 10^{-2}$, near the purely gravitational particle production regime. 
We limit our focus to light bare masses, $10^{-3}< m_{\chi}/H_{\rm end} < 1$, and the results are extrapolated to smaller masses. Following the end of inflation, the dark matter scalar is not yet fully decoupled, and the PSD relaxes to its final form as reheating proceeds. This relaxation is modulated by the oscillations of the inflaton, and is dependent on the dark matter mass $m_{\chi}$, with lighter DM asymptoting to the final value at a smaller rate. This need for a longer evaluation time is in conflict with the increased precision needed to enforce
the Wronskian constraint for $X_k$, and better results are achieved with extrapolation (see Ref.~\cite{Garcia:2022vwm} for further details).
%%%%%%%%%%%%%%%%%%%%%%%%%%%%%%%%%%%%%%%%%%%%%%%%%%%%%%%%%%%%%%%%%%%%%%%%
\section{Dark matter isocurvature}
\label{sec:isocurvature}
%%%%%%%%%%%%%%%%%%%%%%%%%%%%%%%%%%%%%%%%%%%%%%%%%%%%%%%%%%%%%%%%%%%%%%%%
In the previous section, we numerically computed the PSD for the scalar dark matter field $\chi$. The results show that the PSD is red-tilted and has a large amplitude at CMB scales if $\sigma/\lambda < 10^{-1}$. As previously stated, we assume no initial misalignment for the DM scalar field (e.g. $\langle \chi\rangle =0$ and $\langle \chi^2\rangle=0$ initially). As inflation proceeds $\langle \chi\rangle =0$ persists~\cite{Starobinsky:1994bd}, but a non-vanishing value $\langle \chi^2\rangle\neq 0$ is induced from excitations of infrared (superhorizon) modes. This efficient mode excitation during inflation is expected to affect the primordial power spectrum of scalar fluctuations.\footnote{See Refs.~\cite{Ema3, Tenkanen:2019aij} for additional discussions on isocurvature perturbations induced by the long-wavelength (superhorizon) modes.}For a vanishing background value of $\chi$, the fluctuations of the dark matter field will not directly source the curvature perturbation. Therefore, they can be considered to be pure isocurvature fluctuations in the comoving gauge~\cite{Chung:2011xd,Chung:2004nh}. Isocurvature fluctuations have not yet been detected at CMB scales, which imposes an upper bound on the primordial power spectrum of $\mathcal{P}_{\mathcal{S}}(k_*)\lesssim 8.3 \times 10^{-11}$~\cite{Planck:2018jri}.
\begin{figure*}[!t]
\centering
    \includegraphics[width=0.95\textwidth]{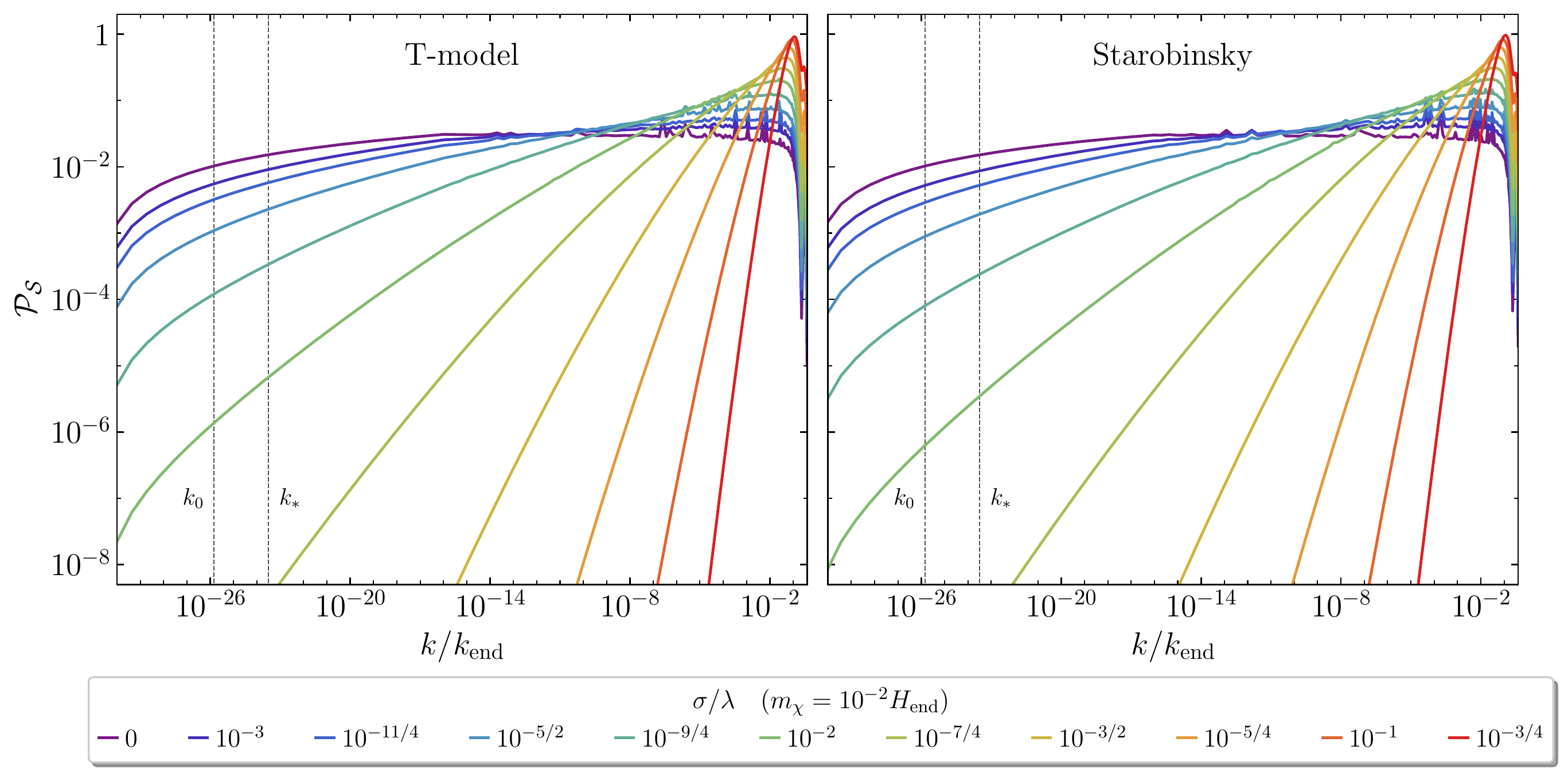}
    \caption{DM isocurvature power spectrum for different inflaton-DM couplings with $m_{\chi}/H_{\rm end}=10^{-2}$, with each coupling represented by a different color. The vertical lines indicate the present horizon scale and the {\em Planck} pivot scale, as shown in Fig.~\ref{fig:PSDs}. Here we set $N_{\rm tot}=76.5$. The left panel corresponds to the T-model, while the right panel corresponds to the Starobinsky model, where the DM coupling $\sigma$ has been rescaled using Eq.~(\ref{eq:sigmaH}). }
    \label{fig:PS_s}
\end{figure*}

The suppression of the linear isocurvature power spectrum results from the geodesic evolution of the background in the scalar field manifold. However, the growth in the energy density of DM is dominated by the quadratic fluctuations that source the variance $\langle \chi^2\rangle$~\cite{Chung:2004nh,Chung:2015pga,Ling:2021zlj,Redi:2022zkt}. The second-order contribution to the isocurvature power spectrum can be approximated using the equation~\cite{Liddle:1999pr,Chung:2004nh,Ling:2021zlj}
\beq 
\label{eq:PS}
\mathcal{P}_{\mathcal{S}}(k) \;=\; \frac{k^3}{2\pi^2\rho_{\chi}^2}\int \diff^3\bx \ \langle \delta\rho_{\chi}(\bx)\delta\rho_{\chi}(0) \rangle e^{-i \bk\cdot\bx}\,,
\eeq
where $\rho_{\chi}$ and $\delta\rho_{\chi}$ denote the DM energy density and its fluctuation, respectively. Upon Fourier transformation, the spectrum can be evaluated in terms of the $\chi$ mode functions and their (conformal) time derivatives~\cite{Ling:2021zlj}. The details of the evaluation of the integral in equation (\ref{eq:PS}) are provided in Appendix~\ref{app:A}. Like the PSD, the late-time isocurvature power spectrum is to be evaluated after DM decoupling. For $10^{-3}\lesssim m_{\chi}/H_{\rm end}$, it is sufficient to evolve the mode functions $X_k$ a few ($\sim 5-6$) $e$-folds after the end of inflation to obtain a good approximation of the final asymptotic value of $\mathcal{P}_{\mathcal{S}}$. 
\par \medskip

\noindent
\textbf{Numerical approach.} 
Following the method detailed in Appendix~\ref{app:A}, we compute the isocurvature power spectrum for various DM masses and inflaton-DM couplings, down to CMB scales and below. Fig.~\ref{fig:PS_s} shows the shape of the isocurvature power spectrum for a broad range of comoving scales with fixed $m_{\chi}/H_{\rm end}=10^{-2}$ and varying coupling ratio $\sigma/\lambda$. The scales are normalized to the comoving momentum of the mode that exited the horizon at the end of inflation. The figure demonstrates that a weak coupling between the inflaton and the dark sector produces a nearly scale-invariant isocurvature spectrum, whose amplitude is in strong disagreement with the {\em Planck} constraint. As the coupling strength increases, the amplitude is suppressed for long wavelength modes, which is consistent with a more efficient particle production at the onset of reheating.

\begin{figure}[!t]
\centering
    \includegraphics[width=\columnwidth]{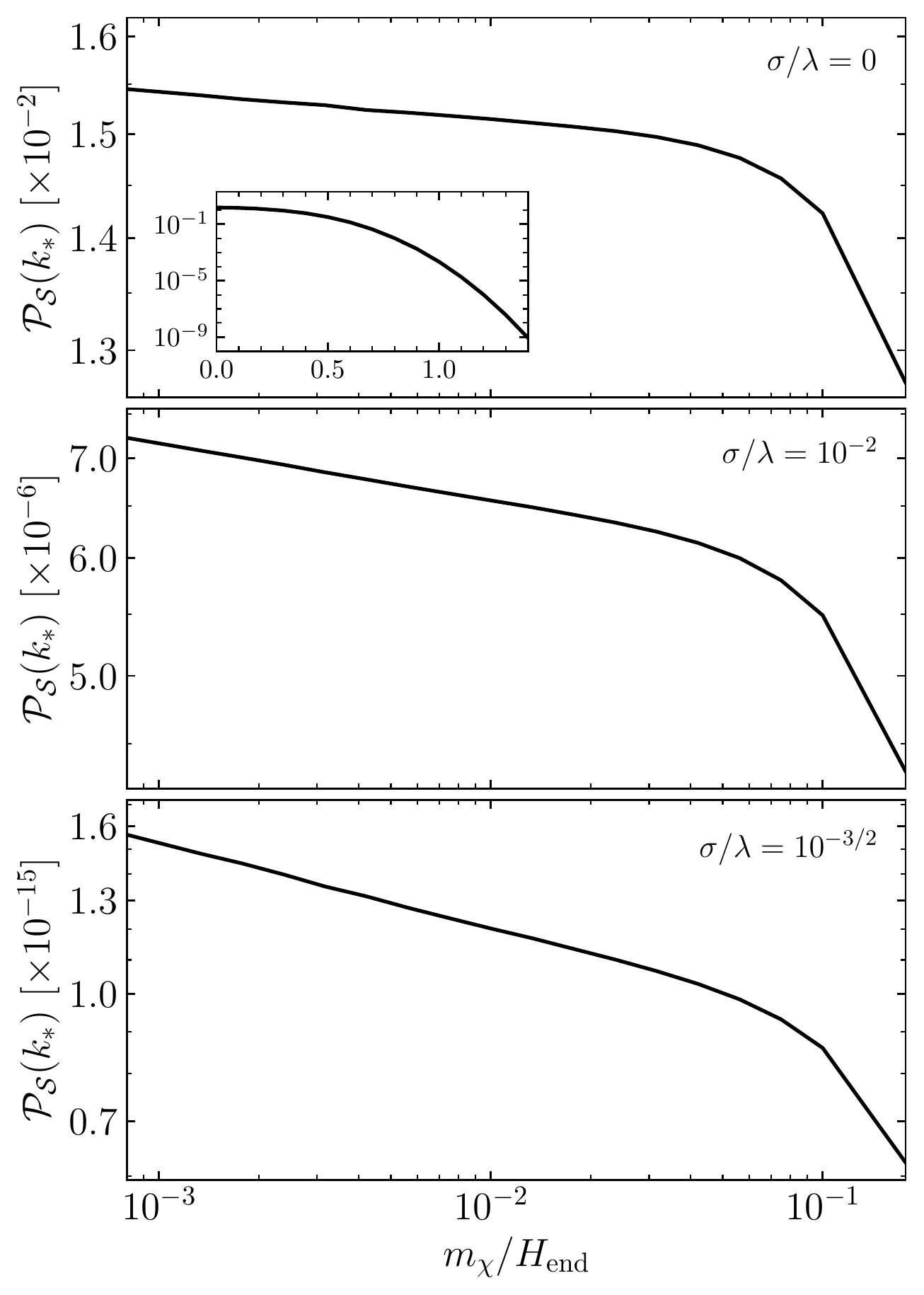}
    \caption{DM isocurvature power spectrum at the {\em Planck} pivot scale $k_*=0.05\,{\rm Mpc}^{-1}$ with $N_*=55$ $e$-folds as a function of the DM mass $m_{\chi}/H_{\rm{end}}$. We show three different inflaton-DM effective couplings $\sigma/\lambda = 0, 10^{-2}$, and $10^{-3/2}$.}
    \label{fig:PS_m}
\end{figure}

Fig.~\ref{fig:PS_m} illustrates the dependence of $\mathcal{P}_{\mathcal{S}}(k_*)$ on the bare DM mass $m_{\chi}$ for three effective couplings $\sigma/\lambda$. We consider the T-model of inflation in this example, but identical constraints can be obtained for the Starobinsky model using the rescaling relationship (\ref{eq:sigmaH}). Importantly, our results are not limited to these specific inflation models and can be applied to different single-field inflationary models in which the dark matter and inflaton fields couple minimally to gravity, by introducing a similar rescaling relationship. As expected, the amplitude of the isocurvature spectrum is smaller for heavy DM since the tachyonic enhancement is weaker (see Eq.~(\ref{eq:omegaH})). Conversely, for $m_{\chi}\ll H_{\rm end}$, a clear red-tilted power-law trend can be observed. Therefore, for a given coupling, we can extrapolate to smaller masses to obtain the full constraints on the minimally coupled scalar DM scenario. We find numerically that for $N_*=55$, the isocurvature constraints are satisfied in the purely gravitational regime when\footnote{We note that the value of the Hubble parameter at the end of inflation, $H_{\rm{end}}$, depends on the specific inflationary potential being considered.}
\begin{align}
    \label{eq:numconstraints1}
    &m_{\chi} \gtrsim 1.11\, H_{\rm{end}} \qquad  (\text{Starobinsky})\, , ~\notag \\
    &m_{\chi} \gtrsim 1.34\, H_{\rm{end}} \qquad (\text{T-model}) \, ,
\end{align}
and in the small bare mass limit $m_{\chi}/H_{\rm end}\ll 1$, we find
\begin{align}
    \label{eq:numconstraints2}
    &\frac{\sigma}{\lambda} \;\gtrsim\; 0.008\qquad  (\text{Starobinsky})\, , ~\notag \\
    &\frac{\sigma}{\lambda} \;\gtrsim\; 0.02 \qquad (\text{T-model}) \, .
\end{align}
ignoring a weak DM mass dependence (see Fig.~\ref{fig:bounds} below). The isocurvature constraints are also weakly sensitive to the value of $N_*$, which in turn implies that they are insensitive to the reheating mechanism. Therefore, these constraints can be expressed in terms of the Hubble parameter at horizon crossing, $H_*$, given by Eqs.~(\ref{eq:isocurvaturecons1}) and (\ref{eq:isocurvatureboundconclusion}). 

\par \medskip

\noindent
\textbf{Analytical approach.}
Appendix~\ref{app:B} explores an analytical derivation of the isocurvature constraint based on the evaluation of the mode functions after inflation. We approximate this evaluation using a finite-duration pure de Sitter era. For the purely gravitational case ($\sigma =0)$, we find that the limit for the Starobinsky model (T-model) is
$m_{\chi} \gtrsim 1.18 H_{\rm{end}} \,(1.41 H_{\rm{end}})$, and for negligible bare mass, we find the limit $\sigma/\lambda \gtrsim 0.006 \, (0.02)$ for the Starobinsky model (T-model) (see Eqs.~(\ref{eq:limpurgrav}) and (\ref{eq:naiveisocurvaturesigmaoverlambda}) and their corresponding derivation). These analytical results are in good agreement with the fully numerical constraints of Eqs.~(\ref{eq:numconstraints1}) and (\ref{eq:numconstraints2}).

However, small discrepancies arise due to several factors. First, some parameters that we approximate as constant actually vary with time as the dynamics of the universe departs from a pure de Sitter era due to slow-roll corrections. In particular, we neglected the explicit time derivative terms for the mode functions in the integrand of the isocurvature expression of Eq.~(\ref{eq:isocurvatureanalytical}). Additionally, the field excursion during inflation is quite large, with $\Delta \phi/M_p \equiv( \phi_*-\phi_\text{end})/M_p \sim  4.7 \,(5.3)$ for the Starobinsky model (T-model), and the end of inflation deviates significantly from a pure de Sitter phase. Typically, for couplings $\sigma/\lambda \sim \mathcal{O}(10^{-2})$, we find that the parameter $\nu^2 = 9/4 - m_{\rm{eff}}^2/H^2$ changes sign between the beginning and end of inflation. This means that the frequency for IR modes switches from a purely imaginary to a real quantity during inflation, making any estimate based on a constant $\nu$ inaccurate. Furthermore, the mode functions continue to evolve after the end of inflation, typically until $a/a_\text{end} \simeq 20-30$, as the universe transitions from a quasi-de Sitter to a matter domination phase.
\par \medskip

\noindent
{\textbf{Long-wavelength modes.}}
Even though we assume at the beginning of inflation that $\langle 0|\chi|0\rangle=0$, for small masses $m_\chi \ll H$ and couplings $\sigma \ll \lambda$, super-horizon modes $k_\text{IR}<k<aH$ are significantly excited as expansion proceeds. In a given Hubble patch, contributions from such modes can be regarded as quasi-homogeneous. One can define a time-dependent effective field value $\chi_\text{hom}\equiv \sqrt{\langle \chi^2 \rangle}$ in such a way that contribution from long wavelengths to the energy density is $\rho_{\chi}\simeq  \frac{1}{2}m_{\chi}^2 \langle \chi^2 \rangle \simeq \frac{1}{2}m_{\chi}^2 \chi_\text{hom}^2$, where $\chi_\text{hom}$ behaves like a coherently oscillating field when $m_\chi \sim H$. The superhorizon contributions to the two-point function of $\chi$ is given as a UV-regularized form~[47]
\begin{equation}
\langle \chi^2 \rangle \;=\; \frac{1}{(2\pi)^3a^2}\int d^3\boldsymbol{k}\, \left( |X_k|^2 - \frac{1}{2\omega_k} \right) \,,
\end{equation}
Nevertheless, it underestimates the total energy density of the dark matter field $\chi$, and does not correctly account for the depletion of this quasi-homogeneous component. Long-wavelength modes continuously re-enter the horizon after inflation, becoming particle-like. These dynamics are accounted for in our analysis, based on individually tracking each field momentum mode via the phase space distribution. Dependence on the IR cutoff $k_\text{IR}$ is discussed in the following paragraph.

\par \medskip

\noindent
\textbf{Infrared sensitivity.} 
As discussed in Appendix~\ref{app:B}, the analytical approach allows us to identify the dependence of our results on the infrared cutoff, $k_\text{IR}$. From Eq.~(\ref{eq:approximatedisocurvature}), we observe that the isocurvature power spectrum 
has a logarithmic dependence on the IR cutoff, given by $\sim \log(k_*/k_\text{IR})$. However, since it is also exponentially sensitive to the ratio of the effective dark matter mass with respect to the Hubble scale during inflation, changing $k_\text{IR}$ by orders of magnitude would only result in a small correction of $\mathcal{O}(1)$ to the isocurvature bound. Therefore, our results are robust and rather insensitive to the infrared cutoff.

%%%%%%%%%%%%%%%%%%%%%%%%%%%%%%%%%%%%%%%%%%%%%%%%%%%%%%%%%%%%%%%%%%%%%%%%
\section{Relic abundance, structure formation, and isocurvature constraints}
\label{sec:relconstraints}
\begin{figure}[!t]
\centering
    \includegraphics[width=\columnwidth]{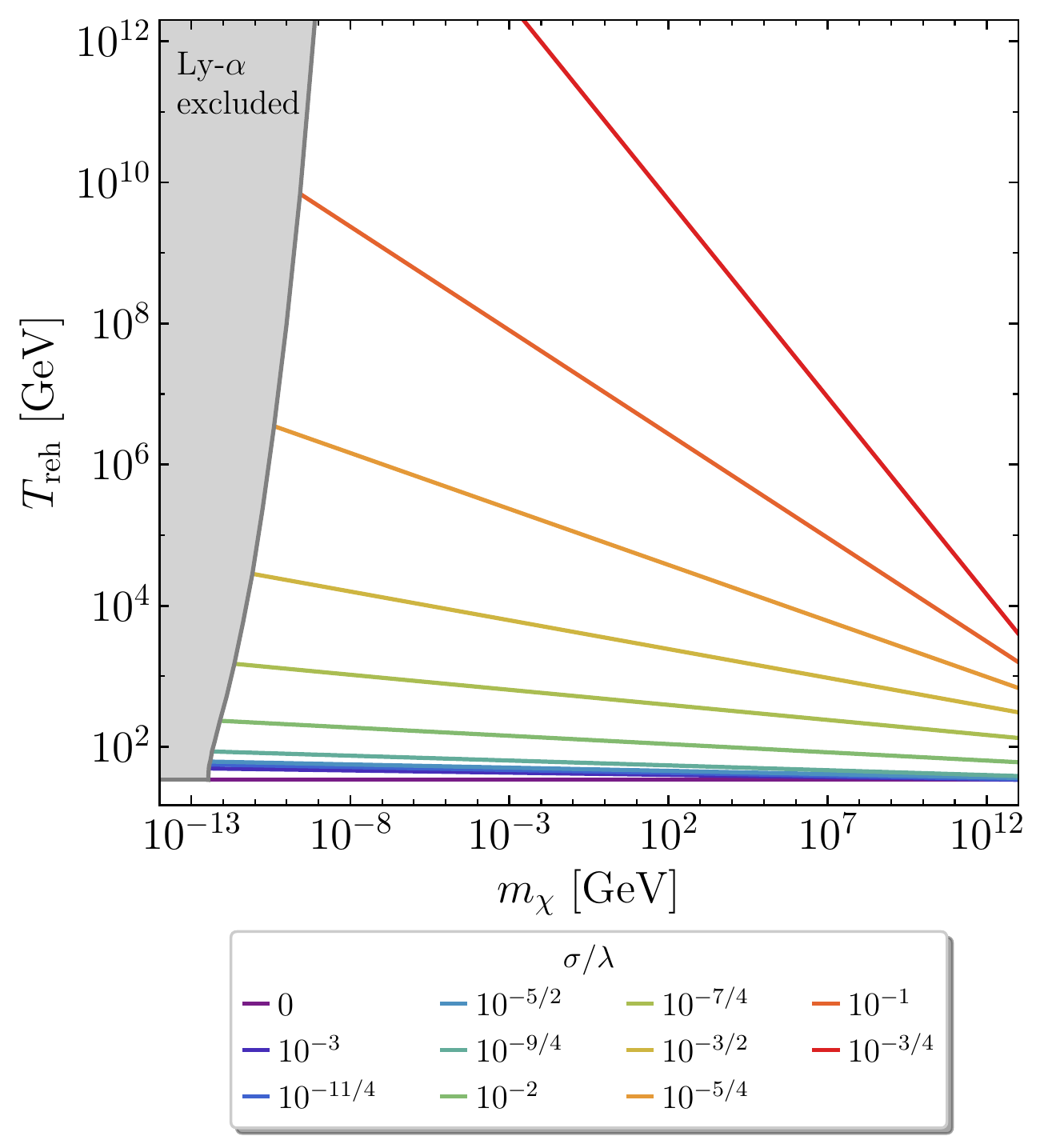}
    \caption{DM masses and reheating temperatures for which the measured DM relic abundance is saturated, for $\sigma/\lambda\ll 1$. The gray shaded region corresponds to the exclusion of light DM masses from the Lyman-$\alpha$ measurement of the matter power spectrum.}
    \label{fig:Omega}
\end{figure}

%%%%%%%%%%%%%%%%%%%%%%%%%%%%%%%%%%%%%%%%%%%%%%%%%%%%%%%%%%%%%%%%%%%%%%%%
We now discuss the implications of the computed phase space distribution (PSD) and isocurvature power spectrum on the model parameter space. First, we must ensure that the measured DM relic abundance, with closure fraction $\Omega_{\rm DM}h^2 = 0.1198$~\cite{Planck:2018vyg}, can be obtained at some reheating temperature $T_{\rm BBN}\leq T_{\rm reh}\leq m_{\phi}$ with $T_{\rm BBN}\sim 1~\text{MeV}$. With the PSD determined, the relic density can be computed as~\cite{Garcia:2022vwm}
\begin{equation}
\Omega_{\rm DM} \;\simeq\; \frac{m_{\chi}n_{\chi}}{\rho_c}=\; \frac{1}{6\pi q_0^3}\left(\frac{m_{\chi} H_0}{M_P^2}\right)\int_{q_0}^{\infty} \diff q\, q^2f_{\chi}(q)\,.
\end{equation}
Here $H_0 = 100 \, h \, \rm{km \, s^{-1} \, Mpc^{-1}}$ is the present Hubble parameter, $\rho_c = 1.05 \times 10^{-5} \, h^2 \, \rm{GeV \, cm^{-3}}$ is the present critical energy density, with $h \sim 0.67$~\cite{Planck:2018vyg}, and we integrate only down to the present comoving scale $q_0$. For the coupling ratio $\sigma/\lambda \ll 1$, saturation of the DM relic density leads to a functional relation between the reheating temperature and the DM mass. This connection was studied in detail in~\cite{Garcia:2022vwm}, and we reproduce the results in Fig.~\ref{fig:Omega}. For pure gravitational production, the relic density is mass-independent, with $\Omega_{\rm DM}h^2\simeq 0.12(T_{\rm reh}/34\,{\rm GeV})$~\cite{Ling:2021zlj,Garcia:2022vwm}. As the inflaton-DM coupling is increased, a functional relation between $T_{\rm reh}$ and $m_{\chi}$ arises, given by $(T_{\rm reh}/1\,{\rm GeV})\propto (m_{\chi}/1\,{\rm GeV})^{\gamma}$, with $-1\leq \gamma\leq 0$. For $\sigma/\lambda\gtrsim 10^{-1/2}$, $\Omega_{\rm DM}\propto T_{\rm reh}m_{\chi}$, corresponding to $\gamma=-1$. Further details can be found in~\cite{Garcia:2022vwm}.

For $\sigma/\lambda \lesssim 10^{-1}$, the DM relic abundance can be saturated with very small DM masses. However, if the DM particles are very light, they could possess a non-negligible pressure component and depart substantially from the cold dark matter paradigm. This departure is manifested as a suppression in the matter power spectrum and the corresponding erasure of clustered structure overdensities. This cutoff in the spectrum is not present down to scales $k_\text{H}(a=1)\simeq 15 \,h\,\text{Mpc}^{-1}$, as determined by the Lyman-$\alpha$ forest measurement of the distribution of matter. For a thermalized DM relic particle (warm dark matter, WDM), this constraint is presented as a lower bound on its mass, $m_\text{WDM}>m_\text{WDM}^{\text{Ly}\mbox{-}\alpha} \simeq (1.9 - 5.3 \, \rm{keV})~at~95\%~{\rm{C.L.}}$~\cite{Narayanan:2000tp,Viel:2005qj,Viel:2013fqw,Baur:2015jsy,Irsic:2017ixq,Palanque-Delabrouille:2019iyz,Garzilli:2019qki}

For out-of-equilibrium DM production mechanisms, the Ly-$\alpha$ bound is dependent on the details of the production and decoupling of the dark particles. Notably, one can translate the WDM lower bound to a bound on the not-equilibrated DM by matching their equation of state parameters. This matching procedure yields the lower bound~\cite{Ballesteros:2020adh}
\beq\label{eq:lyalphaconst}
 \,m_{\chi}^{\text{Ly}\mbox{-}\alpha} \;=\; m_{\rm WDM}^{\text{Ly}\mbox{-}\alpha} \left(\frac{T_{\star}}{T_{\rm WDM,0}}\right)\sqrt{\frac{\langle q^2\rangle}{\langle q^2\rangle_{\rm WDM}}}\,,
\eeq
where $T_{\rm WDM,0}$ is the WDM temperature, $T_{\star}$ is the characteristic energy scale of the produced DM, $T_{\star}=m_{\phi}(a_{\rm end}/a_0)$ for the model discussed in this work, and 
\begin{equation}
    \langle q^2\rangle \, \equiv \, \dfrac{\int \diff q\, q^4 f_\chi(q) }{\int \diff q\, q^2 f_\chi(q) } \,.
    \label{eq:secondmoment}
\end{equation} 
Given the PSD for $\chi$, displayed in Fig.~\ref{fig:PSDs}, we can readily examine the structure formation constraint. However, for weak inflaton-DM coupling, $\langle q^2\rangle$ is dependent on $m_{\chi}$. Therefore, Eq.~(\ref{eq:lyalphaconst}) is a non-linear equation in $m_{\chi}$ whose solution provides the lower bound $m_{\chi}^{\text{Ly}\mbox{-}\alpha}$. For further details, see~\cite{Ballesteros:2020adh, Garcia:2022vwm}.

The Lyman-$\alpha$ constraint is shown in Fig.~\ref{fig:Omega},  leading to the exclusion of masses smaller than $m_{\chi}\simeq 0.34\,{\rm meV}$ for pure gravitational production.
Note that for very small couplings, the lower bound on the DM mass stems from the Ly-$\alpha$ result, whereas for the largest coupling, it is determined by the saturation of $\Omega_{\rm DM}$. 

\begin{figure*}[!t]
\centering
    \includegraphics[width=\textwidth]{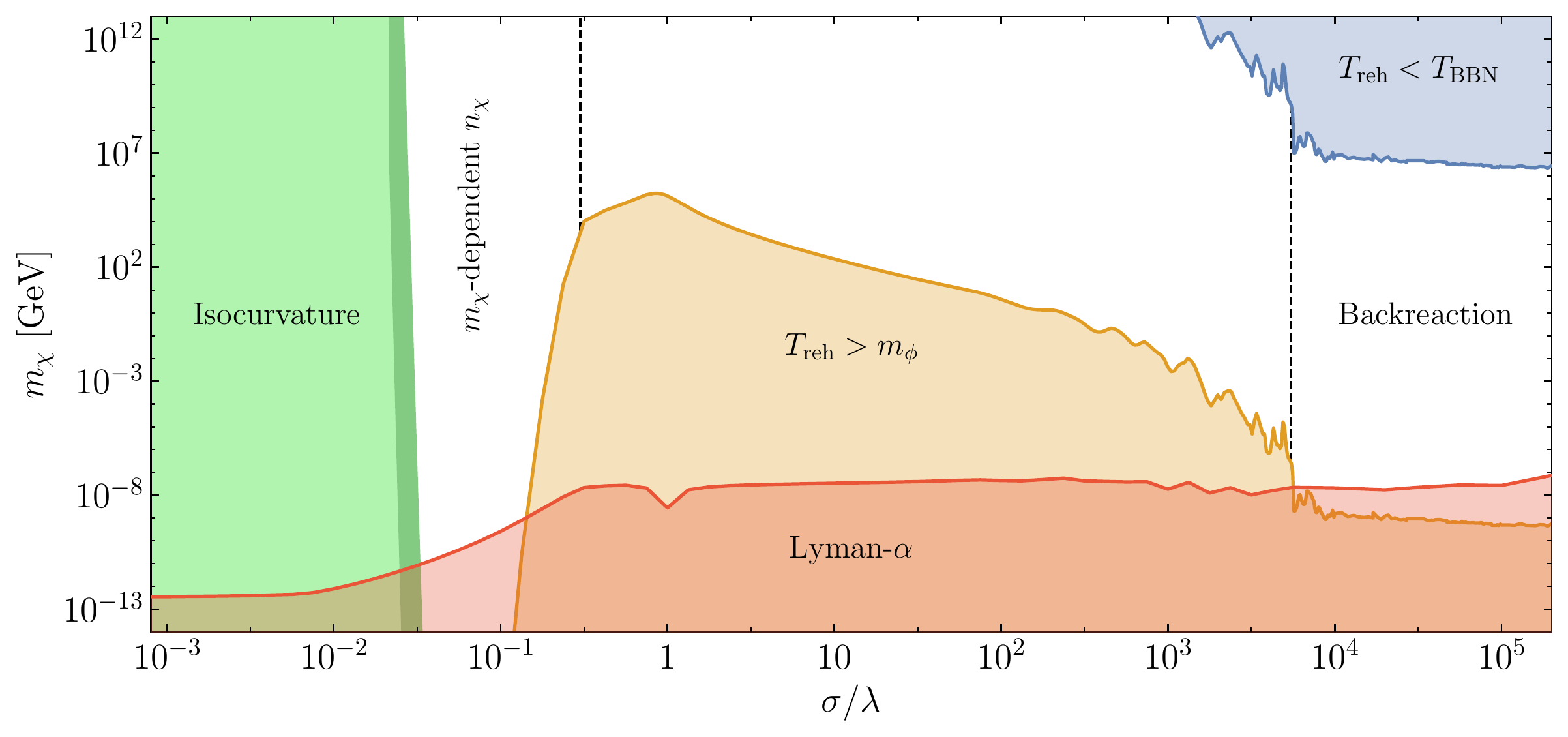}
    \caption{The allowed parameter space (white) for the DM mass and its coupling to the inflaton, in which the relic abundance constraint can be saturated. Forbidden regions correspond to the overproduction of isocurvature (green), the oversuppression of small structure (red), the constraint for a reheating temperature below that required for successful Big Bang Nucleosynthesis (blue), or the constraint for a reheating temperature above what is allowed by perturbative reheating (orange). The width of the boundary of the isocurvature constraint corresponds to a number of $e$-folds $47\leq N_*\leq 55$. The leftmost part of the allowed space corresponds to a region where the power spectrum is sensitive to the DM mass for $m_{\chi}\ll H_{\rm end}$, while the rightmost region corresponds to particle production via broad resonance, resulting in a fragmented inflaton field (see~\cite{Garcia:2022vwm} for details). }
    \label{fig:bounds}
\end{figure*}

The last constraint to be satisfied, and the main result of the present work, corresponds to the bound on the parameter $\beta_{\rm iso}$. This constraint depends on the mass of the dark matter particle $m_{\chi}$, the coupling ratio $\sigma/\lambda$, and the ratio $k_{*}/k_{\rm end}$ (see Figs.~\ref{fig:PS_s} and \ref{fig:PS_m}). Fig.~\ref{fig:PS_m} illustrates that for $\sigma/\lambda<10^{-2}$ and $m_{\chi}\ll H_{\rm end}$, only super-heavy scalar dark matter can avoid the {\em Planck} bound on $\beta_{\rm iso}$~\cite{Ling:2021zlj}. For larger couplings, the allowed parameter space is determined by simultaneously satisfying the constraints on the relic abundance, structure formation, and isocurvature perturbations, and we show it for T-model in Fig.~\ref{fig:bounds}. The red shaded region is excluded by the Ly-$\alpha$ bound. The orange (blue) region is excluded by the requirement that the dark matter density parameter $\Omega_{\rm DM}h^2\simeq 0.12$, subject to the additional constraint $T_{\rm reh}< m_{\phi}$ (and imposing $T_{\rm BBN}<T_{\rm reh}$). These assumptions are necessary to ensure that the post-inflationary reheating can be described by perturbation theory, without conflicting with the predictions of BBN. We note that, for $10^2\lesssim \sigma/\lambda \lesssim 5\times 10^3$, a series of wiggles appear in the $T_{\rm reh}$ exclusion regions. Their presence originates from the quasistochastic nature of the parametric resonance, the dominant DM production mechanism in this regime. The number of broad resonance bands crossed by the DM modes, and the permanence time of the modes in them, do not scale monotonically with the coupling~\cite{Kofman:1997yn}. For $\sigma/\lambda\gtrsim 5\times 10^3$, mode-mode couplings are relevant, driving the inflaton-DM system into the backreaction regime, smoothing out the distributions for the fluctuations, and consequently the shape of the exclusion regions in Fig.~\ref{fig:bounds}.

The light green region in Fig.~\ref{fig:bounds} corresponds to the masses and couplings excluded by the upper bound on $\mathcal{P}_{\mathcal{S}}(k_*)$ for $N_*$. As discussed above, the constraints cannot be satisfied for weak couplings when $m_{\chi}\ll H_{\rm end}$, particularly for pure gravitational production, which is therefore completely excluded. We note the weak dependence of the exclusion region on $m_{\chi}$, which is primarily controlled by the inflaton-DM coupling strength. The combination of Ly-$\alpha$ and $\mathcal{P}_{\mathcal{S}}(k_*)$ limits result in a  lightest allowed DM mass $m_{\chi}\simeq 5\,{\rm meV}$ for $N_*=55$ and $\sigma/\lambda\simeq 0.024$. 

It is worth noting that the green strip in the forbidden domain has a sizable width. This width corresponds to the dependence of the isocurvature spectrum at the {\em Planck} pivot scale on the post-inflationary dynamics, specifically, the relation between $k_*$ and $N_*$. For T-model of inflation, the {\em Planck} constraint on the curvature spectral tilt $n_s$ and the tensor-to-scalar ratio $r$ at the $2\sigma$ level imposes the lower bound $N_*\gtrsim 47$ for the number of $e$-folds after horizon crossing of the pivot scale. On the other hand, $N_*=55$ is close to the perturbative limit~\cite{Ellis:2021kad}. The width of the green strip corresponds then to the saturation of $\beta_{\rm iso}$ between 55 and 47 $e$-folds, with the latter corresponding to the rightmost part of this boundary.
%%%%%%%%%%%%%%%%%%%%%%%%%%%%%%%%%%%%%%%%%%%%%%%%%%%%%%%%%%%%
\section{Discussion and Conclusions}
\label{sec:conclusions}
%%%%%%%%%%%%%%%%%%%%%%%%%%%%%%%%%%%%%%%%%%%%%%%%%%%%%%%%%%%%
The hypothesis of an independent population of the visible and invisible sectors of the universe, arising from the decay of the inflaton field during and after inflation, is not only consistent with the lack of a direct detection signal for DM in the absence of a non-gravitational direct interaction between the Standard Model and dark matter, but also a plausible scenario. In our previous study~\cite{Garcia:2022vwm}, we determined the range of inflaton-DM couplings and DM masses for which the measured DM relic abundance can be produced from inflaton decay, that accounts for the observed DM relic abundance from inflaton decay, while also avoiding the constraint associated with the non-observation of structure formation suppression at small scales. In this study, we further narrow down the available parameter space by taking into account the observation of pure adiabatic fluctuations at CMB scales.
\par \medskip

\noindent
 \textbf{Pure gravitational production.} In the absence of a direct inflaton-dark matter coupling, i.e. $\sigma=0$, the isocurvature power spectrum is nearly scale-invariant and must satisfy the constraint $\mathcal{P}_{\mathcal{S}}(k_*)\lesssim 8.3\times 10^{-11}$ at the {\em Planck} pivot scale. The constraint (\ref{eq:numconstraints1}) can be directly translated to the following bound
\begin{equation}
    \label{eq:isocurvaturecons1}
    m_\chi \, \gtrsim \, 0.54  \, H_* \, ,~~~~~~[\sigma=0]
\end{equation}
where $H_*$ is the Hubble parameter evaluated at the CMB crossing scale. Our results are consistent with previous studies in the pure gravitational regime for $m_{\chi}< H_{*}$~\cite{Chung:2004nh, Chung:2011xd, Ling:2021zlj, Redi:2022zkt}. If we apply this result to the Starobinsky model (T-model), we find the constraint
\begin{equation}
       m_\chi \, > \,  8.5 \, (8.4)   \times 10^{12}~\text{GeV} \, .~~~~~~[\sigma=0]\,.
\end{equation}
\par \medskip

\noindent
 \textbf{Light mass limit.} Furthermore, we explore the form and amplitude of the spectrum in full numerical detail for non-vanishing couplings. In the limit where the bare dark matter mass is much smaller than the Hubble scale, $m_\chi \ll H$, we found that for small couplings, the moderation of the tachyonic growth of DM due to the large field value of the inflaton is inadequate to effectively suppress the isocurvature spectrum for long-wavelength modes. This results in a constraint on the direct coupling given by
\begin{equation}
    \sigma \gtrsim (0.01-0.02) \left( \frac{H_*}{M_P} \right)^2 \, ,
    \label{eq:isocurvatureboundconclusion}
\end{equation}
where the range comes from a weak $m_{\chi}$ and $N_*$-dependence. In terms of the dimensionless ratio $\sigma/\lambda$, for the Starobinsky and T-models, this bound is given by Eq.~(\ref{eq:numconstraints2}). It is worth nothing that these constraints still leave open a range of small masses/couplings where scalar DM can be as light as $5 \, {\rm meV}$ with couplings of $\sigma/\lambda \sim \mathcal{O}(10^{-1})$. Smaller masses and/or couplings might be achievable in scenarios with non-minimal conformal coupling, and we intend to explore such models in detail in future work. 
\par \medskip

\noindent
\textbf{Generality and scope of our results.} We have introduced an infrared cutoff $k_\text{IR}$ that selects only modes corresponding to wavelengths of the order of maximal causal distances at present. Our numerical results, supplemented by analytical estimates, have demonstrated that the isocurvature bound derived in this paper exhibits a logarithmic dependence on the infrared cutoff, but an exponential dependence on the effective dark matter mass during inflation. As a result, our bounds are only weakly dependent on the IR cutoff. Moreover, we have shown that our findings for the Starobinsky and T-model of inflation can be translated to one another by rescaling the parameter $\lambda$, associated with the Hubble scale at the end of inflation, $H_{\rm{end}}$. However, we note that the isocurvature bound depends on the Hubble parameter at horizon exit, $H_*$, as shown by Eqs.~(\ref{eq:isocurvaturecons1}) and~(\ref{eq:isocurvatureboundconclusion}). Therefore, the isocurvature bound derived in this study can be extended to a wider range of plateau-like single-field inflationary potentials, where the inflaton and dark matter fields couple minimally to gravity and have different values of $H_*$.

Looking towards the future, upcoming experiments such as CMB-S4~\cite{Abazajian:2019eic} and LiteBIRD~\cite{Hazumi:2019lys} will not only be geared towards the detection of B-modes in the CMB, but will also offer a more comprehensive analysis of the CMB scalar power spectrum. This will lead to stronger constraints on the scalar tilt, $n_s$, and tensor-to-scalar ratio, $r$, paving the way for a new era of precision cosmology. Moreover, the improved analysis is also expected to strengthen the constraints on isocurvature modes. If no isocurvature modes are detected by these experiments, the resulting bound will narrow down the parameter space for gravitationally produced dark matter. Such an outcome will be of paramount importance for future dark matter searches and will undoubtedly provide important insights about the nature of dark matter.

\begin{acknowledgments}

\noindent
We would like to thank Mustafa Amin, Veronica Guidetti, Andrew Long, Michele Redi, and Sai Chaitanya Tadepalli for helpful discussions. MG is supported by the DGAPA-PAPIIT grant IA103123 at UNAM. MP acknowledges support by the Deutsche Forschungsgemeinschaft (DFG, German Research Foundation) under Germany's Excellence Strategy – EXC 2121 “Quantum Universe” – 390833306. The work of S.V. was supported in part by DOE grant DE-SC0022148. This work was made possible by the support of the Institut Pascal at Université Paris-Saclay during the Paris-Saclay Astroparticle Symposium 2022, with the support of the P2IO Laboratory of Excellence (program “Investissements d’avenir” ANR-11-IDEX-0003-01 Paris-Saclay and ANR-10-LABX-0038), the P2I axis of the Graduate School Physics of Université Paris-Saclay, as well as IJCLab, CEA, IPhT, APPEC, the IN2P3 master projet UCMN and EuCAPT ANR-11-IDEX-0003-01 Paris-Saclay and ANR-10-LABX-0038). Numerical results were partially obtained from a custom Fortran code utilizing the thread-safe arbitrary precision package MPFUN-For~\cite{mpfun}, at the Holiday cluster at Instituto de F\'isica, UNAM.
\end{acknowledgments}

\appendix

%%%%%%%%%%%%%%%%%%%%%%%%%%%%%%%%%%%%%%%%%%%%%%%%%%
\section{Quadratic Isocurvature}
\label{app:A}
%%%%%%%%%%%%%%%%%%%%%%%%%%%%%%%%%%%%%%%%%%%%%%%%%%
In this appendix, we show how to evaluate the power spectrum of the DM isocurvature fluctuation at second order in perturbation theory. As mentioned in Sec.~\ref{sec:isocurvature}, in the absence of a DM misalignment, it may be evaluated as~\cite{Liddle:1999pr,Chung:2004nh,Ling:2021zlj}
\begin{align}
\mathcal{P}_{\mathcal{S}}(k) \;&=\; \frac{1}{\rho_{\chi}^2} \frac{k^3}{2\pi^2}\int \diff^3\bx \ \langle \delta\rho_{\chi}(\bx)\delta\rho_{\chi}(0) \rangle e^{-i \bk\cdot\bx}\\ \label{eq:Psp2}
&=\; \frac{k^3}{(2\pi)^5\rho_{\chi}^2a^8}\int \diff^3\bp \ P_X(p,|\bp-\bk|)\,,
\end{align}
where the integrand in the second line is defined as
\begin{align}\notag
P_X(p,q)\;&=\; |X'_p|^2|X'_{q}|^2 + a^4m_{\chi}^4 |X_p|^2|X_{q}|^2\\
& + a^2m_{\chi}^2\left[ (X_pX_p^{\prime *})(X_{q}X_{q}^{\prime *}) + {\rm h.c.} \right]\,.
\label{eq:integrandisocurvature}
\end{align}
The integral (\ref{eq:Psp2}) needs to be treated carefully. When $m_{\chi}\ll H_{\rm end}$ and $\sigma/\lambda \ll 1$, $P_X(p,q)$ is divergent in the IR for both arguments. However, this divergence can be naturally regularized by imposing a lower momentum (IR) cutoff. As discussed in the main text, this cutoff scale can be associated with the mode that left the horizon at the beginning of inflation (our choice for $\mathcal{P}_{\mathcal{S}}$), or more conservatively, with the wavenumber of the present-day horizon scale. Additionally, it is important to evaluate the power spectrum $\mathcal{P}_{\mathcal{S}}$ after the decoupling time of DM, which in this case corresponds to the end of reheating. In scenarios where the amplitude of isocurvature at CMB scales violates present constraints, the majority of the contribution to the spectrum comes from modes excited during inflation. Hence, it is sufficient to evaluate the spectrum during reheating, but well after the end of inflation.

\begin{figure}[!t]
%\vspace{-10pt}
\centering
\begin{tikzpicture}%
		\fill[Apricot] (1.15,1.15) - - (4.8,4.8) - - (4.8,2.5) - - (2.8,0.5) - - (1.8,0.5) - - cycle;
    		\draw[->, line width=0.6pt] (0,0) -- (6.3,0);%
    		\draw[->, line width=0.6pt] (0,0) -- (0,6.3);%
    		\draw[line width=0.6pt] (0,2.3) -- (2.3,0);%
    		\draw[line width=0.6pt] (0,2.3) -- (3.7,6);%
    		\draw[line width=0.6pt] (2.3,0) -- (6,3.7);%
    		\draw[dashed,line width=0.6pt] (0,0) -- (4.85,4.85);%
    		\draw[dashed,line width=0.6pt] (4.8,0) -- (4.8,6);%
    		\draw[dashed,line width=0.6pt] (0,4.8) -- (6,4.8);%
    		\node at (4.8,-0.4) {$k_{\rm UV}$};
    		\node at (-0.5,4.8) {$k_{\rm UV}$};v
    		\draw[dashed,line width=0.6pt] (0,0.5) -- (6,0.5);%
		\draw[dashed,line width=0.6pt] (0.5,0) -- (0.5,6);%
    		%\node at (-0.5,2.7) {$-\tau_0^{-1}$};
		\node at (-0.5,0.6) {$k_{\rm IR}$};
    		\node at (2.3,-0.3) {$k$};
    		\node at (-0.3,2.3) {$k$};
    		\node at (6.3,-0.4) {$q$};
    		\node at (-0.3,6.3) {$p$};
  \end{tikzpicture}
  \caption{Illustration of the integration domain in Eq.~(\ref{eq:PSp4}), corresponding to the case with $2k_{\rm IR} < k <k_{\rm UV}-k_{\rm IR}$.} \label{fig:region}
\end{figure}
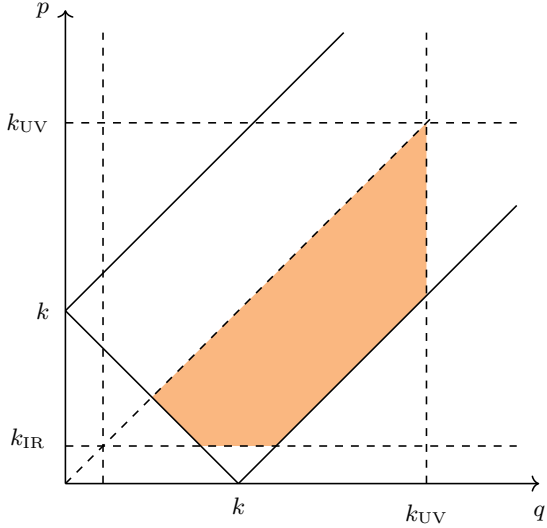

For finite DM masses and couplings, it is necessary to numerically evaluate the spectrum and correctly exclude  the IR ($p\rightarrow 0$) and the collinear ($|\bp-\bk|\rightarrow 0$) divergences from the integral. We find it convenient to use the following integration strategy for Eq.~(\ref{eq:Psp2}),
\begin{align}
\mathcal{P}_{\mathcal{S}}(k) \;&=\; \frac{k^2}{(2\pi)^4\rho_{\chi}^2a^8}\int_0^{\infty} \diff  p\ p \int_{|k-p|}^{k+p}\diff q\ q \ P_X(p,q) \notag \\ 
&=\; \frac{2k^2}{(2\pi)^4\rho_{\chi}^2a^8} \, \mathcal{I} \big(P_X(p,q)\big) \,, \label{eq:PSp3}
\end{align}
with
\vskip-0.2in
\begin{widetext}
\begin{equation}
\mathcal{I} \big(f(p,q)\big) \, \equiv \, \left(\int_{k_{\rm IR}}^{k/2} \diff p \int_{k-p}^{k+p}\diff  q + \int_{k/2}^{k_{\rm UV}-k} \diff p \int_{p}^{k+p}\diff q + \int_{k_{\rm UV}-k}^{k_{\rm UV}} \diff p \int_{p}^{k_{\rm UV}}\diff q\right) pq\,f(p,q)\,.
\label{eq:PSp4}
\end{equation}
\end{widetext}
We have included a UV cutoff for definiteness, which may be identified with the decoupling scale, last scale excited at the end of reheating. However, this scale is not necessary for the convergence of the integral in the UV. The expression (\ref{eq:PSp3}) is valid under the assumption that $2k_{\rm IR} < k <k_{\rm UV}-k_{\rm IR}$. The spectrum for wavenumbers near the boundary values can be determined by identifying the proper integration limits, as shown in Fig.~\ref{fig:region}. \par \medskip

\par \medskip
%%%%%%%%%%%%%%%%%%%%%%%%%%%%%%%%%%%%%%%%%%%%%%%%%%%%%%%%%%%%%%%%%%%%%%
\section{Analytical Derivation of the Isocurvature Power Spectrum}
\label{app:B}
%%%%%%%%%%%%%%%%%%%%%%%%%%%%%%%%%%%%%%%%%%%%%%%%%%%%%%%%%%%%%%%%%%%%%%
In this section, we provide a detailed analytical derivation of the isocurvature power spectrum. We approximate the dynamics of the universe during inflation as a de Sitter phase characterized by a constant accelerated expansion with $a(t) = e^{Ht}$ and a constant Hubble parameter. We assume that this phase begins at some initial time $a(t_i) = a_i$ and ends at $a(t_{\rm{end}}) = a_{\rm{end}}$. To obtain the isocurvature power spectrum, we solve the mode equation for $ \chi_k=  X_k/a$ expressed as a function of cosmic time, which is given by
\begin{equation}
    \ddot{\chi}_k+3 H \dot \chi_k +\left( \dfrac{k^2}{a^2} +m_\chi^2 \right) \chi_k  \, = \, 0 \, ,
\end{equation}
where dots denote derivatives with respect to cosmic time and $ m_\chi$ is the effective dark matter mass, which is assumed to be constant. Assuming Bunch-Davies boundary conditions and a pure de Sitter phase, the solution to the previous equation can be expressed as
\begin{equation}
    \chi_k(t) \, = \, \dfrac{\sqrt{\pi}}{2 a^{3/2}\sqrt{H}} e^{ i \frac{\pi}{2} (\nu+\frac{1}{2})} H_\nu^{(1)} \left(\dfrac{k}{aH}\right) \, ,
\end{equation}
where $H_\nu^{(1)} $ is the Hankel function of the first kind, and
\begin{equation}
    \nu \, \equiv \, \sqrt{\dfrac{9}{4}-\dfrac{m_\chi^2}{H^2}} \, .
\end{equation}
In the large wavelength limit $k \ll aH$, this solution reduces to
\begin{widetext}
\begin{equation}
    \chi_k(k \ll aH) \, \simeq \, \dfrac{1}{2\sqrt{\pi} a^{3/2}\sqrt{H}} e^{ i \frac{\pi}{2} (\nu-\frac{1}{2})} \left( e^{-i \pi \nu}\Gamma(-\nu) \left( \dfrac{k}{2aH} \right)^{\nu} + \Gamma(\nu) \left( \dfrac{k}{2aH} \right)^{-\nu} \right)  \, ,
\end{equation}
\end{widetext}
in agreement with the results from Ref.~\cite{Kolb:2023dzp}. The power spectrum of isocurvature perturbations can be estimated at the end of inflation by neglecting the time derivative of mode functions in Eq.~(\ref{eq:integrandisocurvature}). Then Eq.~(\ref{eq:PSp3}) becomes
\begin{align}
\mathcal{P}_{\mathcal{S}}(k) \;&=\; \frac{k^2}{(2\pi)^4\langle \chi^2 \rangle^2}\int_0^{\infty} \diff  p\ p \int_{|k-p|}^{k+p}\diff q\ q \ |\chi_p|^2 |\chi_q|^2\,,\notag \\ &=\; \frac{2k^2}{(2\pi)^4 \langle \chi^2 \rangle^2}\, \mathcal{I} \big(|\chi_p|^2 |\chi_q|^2 \big) \, ,
\label{eq:isocurvatureanalytical}
\end{align}
where we approximated $\rho_\chi \simeq m_\chi^2 \langle \chi^2 \rangle$, which is valid at later times. The variance of the dark matter field $\chi$ is given by
\begin{equation}
    \langle \chi^2 \rangle \, \equiv \, \int_{k_\text{IR}}^{k_\text{UV}} \, \mathcal{P}_\chi \, \dfrac{\diff k}{k} \, ,
\end{equation}
with the power spectrum defined as 
\begin{equation}
    \mathcal{P_\chi} \, \equiv \, \dfrac{k^3}{2 \pi^2}| \chi_k |^2 \, .
\end{equation}
Here, we introduce long wavelength (IR) and short wavelength (UV) cutoffs to regularize the power spectrum, following the approach outlined in \cite{Vilenkin:1983xp}, and define $k_\text{IR} = a_i H$ and $k_\text{UV}= a_{\rm{end}} H$. We use the notation $N_\text{tot} \equiv \log(a_{\rm{end}}/a_i)$ to represent the total duration of the de Sitter era in terms of $e$-folds, while $\Delta N_* > 0$ denotes the duration of inflation in $e$-folds between the CMB pivot scale $k_*$ crossing and the end of inflation. In what follows, we calculate the power spectra for $\chi$, variances, and the corresponding isocurvature power spectrum evaluated at $k_*$. We consider three different regimes.

\par \medskip
\noindent
\textbf{Negligible mass limit $m_\chi \ll H$.} 
If the DM mass can be neglected compared to the Hubble scale during the de Sitter phase, then $\nu \simeq 3/2$, and the mode function can be expressed as
\begin{equation}
    | \chi_k(k \ll aH) | \, \simeq \, \frac{1}{\sqrt{2} \sqrt{a^3 H}} \left(\frac{k}{a H}\right)^{-3/2} \, = \frac{H}{\sqrt{2} k^{3/2}} \, ,
    \label{eq:modereal}
\end{equation}
which leads to a well-known scale-invariant result for the power spectrum $
\mathcal{P_\chi} \, = \, H^2/(4 \pi^2) $. Combining Eq.~(\ref{eq:modereal}) with Eq.~(\ref{eq:PSDandoccupationnumber}) and neglecting the time derivative of the mode function, we find that in the long wavelength (IR, $k \ll aH$) regime, the phase space distribution $f_{\chi}(k,t)$ scales as $f_{\chi}(k,t)\sim |\chi_k|^2 \sim k^{-3}$. The variance of the field at the end of the de Sitter phase is
\begin{equation}
    \langle \chi^2 \rangle \, \simeq \,  \dfrac{H^2}{4 \pi^2} \log \left( \dfrac{a_{\rm{end}} }{a_i } \right) \, = \,\dfrac{H^2}{4 \pi^2} N_\text{tot}  \, .  
    \label{eq:variancesmallmass}
\end{equation}
 Evaluated for the CMB crossing scale, the isocurvature power spectrum~(\ref{eq:isocurvatureanalytical}) is
\begin{equation}
 \mathcal{P}^{m_\chi \ll H}_{\mathcal{S}}(k_*) \, \simeq \, \dfrac{N_\text{tot}-\Delta N_*}{N_\text{tot}^2} \, ,
 \label{eq:isocurvaturenegligiblemass}
\end{equation}
which scales as $1/N_\text{tot}$ for $\Delta N_* \ll N_\text{tot}$.

\par \medskip

\begin{figure}[!t]
\centering
    \includegraphics[width=\columnwidth]{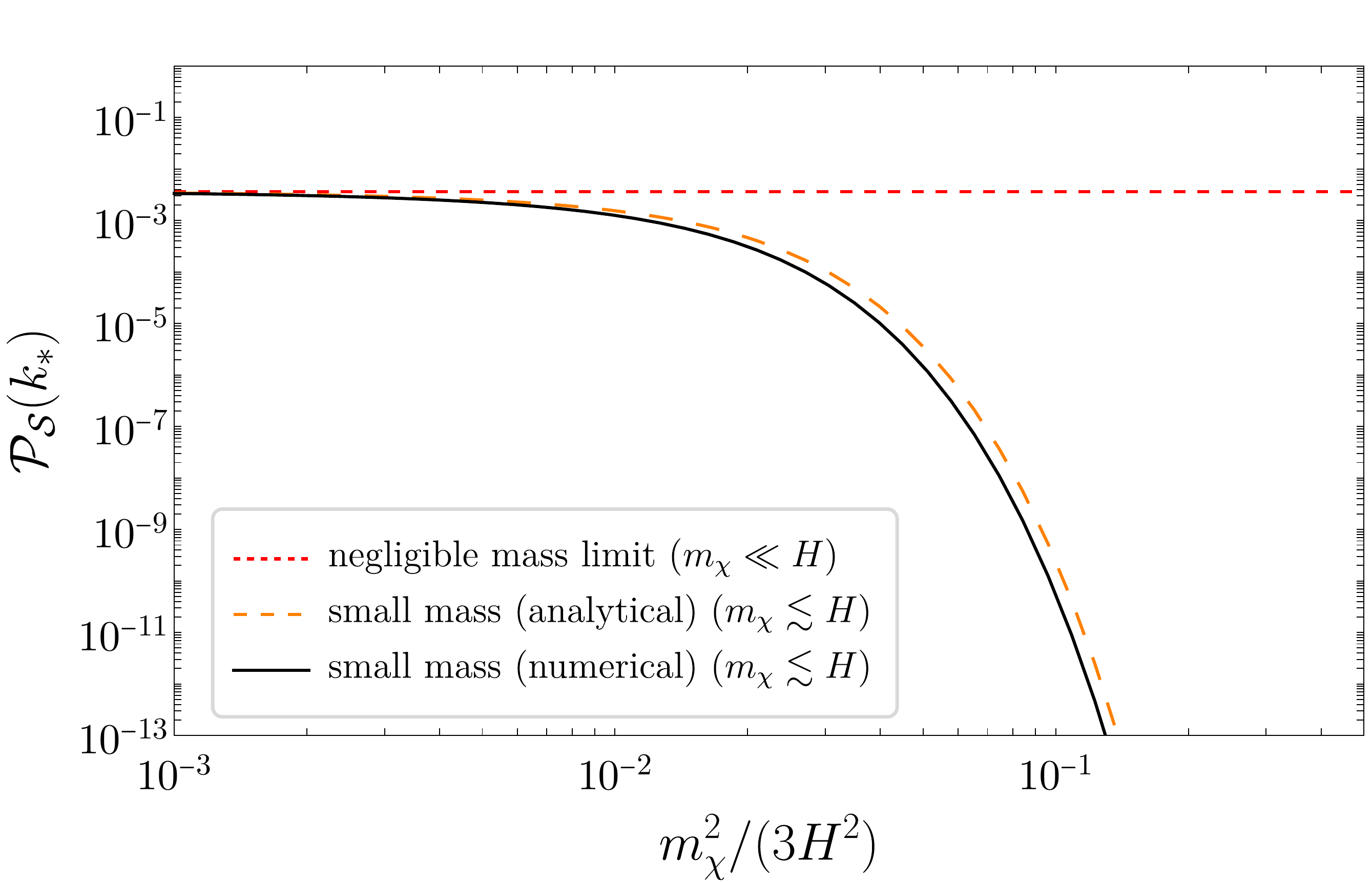}
    \caption{Isocurvature power spectrum as a function of dark matter mass. The red dashed line represents results based on the analytical solution for the mode function in the negligible mass limit of Eq.~(\ref{eq:isocurvaturenegligiblemass}). The orange dashed line represents the approximated expression of Eq.~(\ref{eq:approximatedisocurvature}) in the small mass regime, and the black line a numerical evaluation of Eq.~(\ref{eq:isocurvatureanalytical}) in the same regime for comparison.}
    \label{fig:PS_analytical}
\end{figure}

\noindent
\textbf{Small mass limit $m_\chi \lesssim H$.} If the DM mass is small but not negligible compared to the Hubble scale, $\nu \simeq 3/2 - \beta$ with $\beta\equiv m_\chi^2/(3 H^2)$, then the mode function can be approximated by~\cite{Mukhanov:1990me, Liddle:1999pr}
\begin{equation}
    | \chi_k(k \ll aH) | \, \simeq \,  \dfrac{1}{\sqrt{2}} H (a H)^{-3/2} \left(\frac{k}{a H}\right)^{-\nu } \,,
    \label{eq:moderealmedium}
\end{equation}
which corresponds to a blue-tilted power spectrum
\begin{equation}
    \mathcal{P_\chi} \,= \dfrac{H^2}{4 \pi^2} \left( \dfrac{k}{aH} \right)^{ 2 \beta} \, ,
\end{equation}
when $\beta>0$.
In this regime, the phase space distribution $f_\chi(k,t)$ scales as $f_\chi(k,t)\sim |\chi_k|^2 \sim k^{-2 \nu} \sim k^{2\beta - 3}$. The variance of the field at the end of the de Sitter phase is given by
\begin{equation}
    \langle \chi^2 \rangle \, \simeq \,  \left(\frac{1}{a H}\right)^{2 \beta } \frac{H^2}{8 \pi ^2 \beta }    \left(k_\text{UV}^{2 \beta }-k_\text{IR}^{2 \beta }\right)\, .     \label{eq:variancemediummass}
\end{equation}
If $\beta>10^{-2}$ and $k_\text{IR}= a_i H \simeq 10 ^{-10} k_*$ and $k_\text{UV}= a_{\rm{end}} H \simeq 10^{24} k_*$, the variance becomes insensitive to the IR. In this regime, the isocurvature power spectrum can be expressed as a sum of hypergeometric functions. Alternatively, one can isolate the $\beta$ dependence and derive the following approximate expression
\begin{equation}
     \mathcal{P}^{m_\chi \lesssim H}_{\mathcal{S}}(k_*) \;\simeq \; \dfrac{ \mathcal{P}^{m_\chi \ll H}_{\mathcal{S}}(k_*)}{\mathscr{F}_0} \, \mathscr{F}\left( \dfrac{m_\chi^2}{3 H^2} \right) \,,
     \label{eq:approximatedisocurvature}
\end{equation}
where
\begin{equation}
     \mathscr{F} (x) \, \equiv \, \dfrac{x^2 e^{-4 \Delta N_* x}}{16 \left(1-e^{-2 N_\text{tot} x}\right)^2}\,.
     \label{eq:curlyFfunction}
\end{equation}
and $\mathscr{F}_0=\lim_{x\rightarrow 0}\mathscr{F}(x)=1/(64N_\text{tot}^2)$. This function shows that the isocurvature decays exponentially with the ratio of the DM mass to the Hubble scale, as shown in~\cite{Redi:2022zkt}. Note that the function $\mathscr{F} (x)$ is insensitive to $N_\text{tot}$ in the limit $N_\text{tot} x \gg 1$. Therefore, the only significant infrared sensitivity of Eq.~(\ref{eq:approximatedisocurvature}) arises from the scaling $\mathcal{P}^{m_\chi \ll H}_{\mathcal{S}}(k_*) / \mathscr{F}_0\sim N_\text{tot}$ at large $N_\text{tot}$, which is equivalent to a logarithmic scaling in terms of the infrared cutoff, $\sim \log(k/k_\text{IR})$. A comparison between the Eq.~(\ref{eq:approximatedisocurvature}) and a numerical evaluation of the isocurvature power spectrum of Eq.~(\ref{eq:isocurvatureanalytical}) is shown in Fig.~\ref{fig:PS_analytical}, indicating good agreement between both results.
\par \medskip

\noindent
\textbf{Large mass limit $m_\chi \gtrsim H$.} 
For $m_\chi^2>9 H^2/4$, $\nu \in \mathbb{C}$. We define $\nu \equiv -i \tilde{\nu}$ with $\tilde \nu \in \mathbb{R}^+$ and $\tilde \nu^2 \simeq m_\chi^2 / H^2$. In this case, the mode function can be expressed as
 \begin{widetext}
\begin{equation}
        \chi_k(k \ll aH) \, \simeq \,\dfrac{1}{2\sqrt{\pi} a^{3/2}\sqrt{H}} e^{  \frac{\pi}{2} (-\tilde \nu-\frac{1}{2}i)} \big[   e^{\pi \tilde \nu} \Gamma ( - i \tilde \nu ) e^{i \tilde \nu \log(x/2)}+ \Gamma ( i \tilde \nu ) e^{-i \tilde \nu \log(x/2)}         \big] \, ,
\end{equation}
where $x\equiv k/(aH)$, and
\begin{equation}
     |   \chi_k(k \ll aH) |\, \simeq \,\dfrac{e^{-\pi \tilde \nu /2}}{2\sqrt{\pi} a^{3/2}\sqrt{H}} \, \mathcal{F}(\tilde \nu)\, ,
\end{equation}
where the function $\mathcal{F}(\tilde \nu)$ is defined as
\begin{equation}
    \mathcal{F}(\tilde \nu) \, \equiv \, \left(
    e^{\pi \tilde \nu} \Big[ 2 \cos \big(2 \tilde \nu \log(x/2) \big)  \Big( \mathcal{R}_{\tilde \nu}^2 -  \mathcal{I}_{\tilde \nu}^2 \Big) + 4 \mathcal{R}_{\tilde \nu} \mathcal{I}_{\tilde \nu} \sin \big(2 \tilde \nu \log(x/2) \big) \Big]
    + \left( 1 + e^{2 \pi \tilde \nu} \right) \left( \dfrac{\pi}{\tilde \nu \sinh (\pi \tilde \nu)} \right) \right)^{1/2} \, .
\end{equation}
\end{widetext}
Here $\mathcal{R}_{\tilde \nu}\equiv \text{Re}(\Gamma(i \tilde \nu))$ and $\mathcal{I}_{\tilde \nu}\equiv \text{Im}(\Gamma(i \tilde \nu))$. The first term in square brackets in this expression is an oscillatory term with frequency $2 \tilde \nu \log(k/(2aH))$, which can typically be neglected. This function can be well approximated in the limit of large $\tilde \nu$ by
\begin{equation}
    \mathcal{F}(\tilde \nu \gg 1) \, \simeq \, \sqrt{\dfrac{2 \pi }{\tilde \nu}} e^{\pi \tilde \nu /2} \,.   
\end{equation}
which is a good approximation even for $\tilde \nu \sim \mathcal{O}(1)$. The modulus of the mode function can be approximated as
\begin{equation}
     |   \chi_k(k \ll aH) |\, \simeq \,\dfrac{1}{ a^{3/2}\sqrt{H}} \dfrac{1}{\sqrt{2 \tilde \nu}} \, \simeq \dfrac{1}{ \sqrt{2 m_\chi a^3}} \, ,
    \label{eq:modeimaginary}
\end{equation}
and the power spectrum can be expressed as
\begin{equation}
    \mathcal{P_\chi} \, \simeq \, \dfrac{1}{4 \pi^2 a^3 m_\chi} k^3  \, \simeq  \dfrac{1}{4 \pi^2 m_\chi} H^3 e^{3(N_k-N)}\, \,.
\end{equation}
Here $N_k$ is the number of $e$-folds at which the scale $k$ crosses the horizon $a(N_k)H=k$. The spectrum is well behaved in the IR but decreases exponentially with the number of $e$-folds after the horizon crossing $N-N_k>0$, redshifting as non-relativistic matter. The variance of the field at the end of de Sitter phase is~\cite{Enqvist:1987au}
\begin{equation}
    \langle \chi^2 \rangle \, \simeq  \,  \dfrac{H^3}{12 \pi^2  m_\chi} \, ,
    \label{eq:variancelargemass}
\end{equation}
where we consider $m_{\chi} \sim H$.\footnote{In this appendix, we use the pure de Sitter solutions to the mode equations as an approximation to compute the power spectra expected after a finite quasi-de Sitter phase. It should be noted that when $\nu$ is imaginary, a naive UV cutoff in a pure de Sitter universe may not be sufficient to subtract all the UV divergences, and higher-order adiabatic subtraction terms may be required instead. However, here we only want to illustrate that the isocurvature power spectrum is suppressed and a naive UV cutoff is sufficient for our treatment in the limit $m_{\chi} \sim H$. For a detailed discussion of the optimal order of truncation of the adiabatic expansion, see Refs.~{\cite{Dabrowski:2014ica, Corba:2022ugu}} for a detailed discussion of the optimal order of truncation of the adiabatic expansion.} In this limit, the isocurvature power spectrum is exponentially suppressed:
\begin{equation}
 \mathcal{P}^{m_\chi \gtrsim H}_{\mathcal{S}}(k_*) \;\simeq \;   \frac{3 e^{-3 \Delta N_*}}{2} \, ,
 \label{eq:isocurvaturelargemass}
\end{equation}
and for $\Delta N_*=55$ we find $\mathcal{P}^{m_\chi \gtrsim H}_{\mathcal{S}}(k_*) \simeq \mathcal{O}(10^{-72})$.\\

\noindent
\textbf{Estimating the Isocurvature Constraint.}
From Eq.~(\ref{eq:isocurvaturelargemass}), one can readily see that the isocurvature constraint $\mathcal{P}_{\mathcal{S}}(k_*) < \beta_\text{iso} \mathcal{P}_{\mathcal{R}}(k_*) \simeq  10^{-11}$ is always satisfied for $m_\chi^2>9 H^2/4$. 

In the limit of negligible mass, where $m_{\chi} \ll H$, as described in Eq.~(\ref{eq:isocurvaturenegligiblemass}), the isocurvature power spectrum is too large by orders of magnitude, given by $ \mathcal{P}^{m_\chi \ll H}_{\mathcal{S}}(k_*) \, \sim \, \mathcal{O}(10^{-3})$ for $N_\text{tot} \, \sim \, \mathcal{O}(100)$. 

In the intermediate regime, where $m_\chi \lesssim H$, we can use the approximate expression~(\ref{eq:approximatedisocurvature}) to compute the limit
\begin{equation}
m_\chi\, \gtrsim \, 0.57 \, H \, ,
\label{eq:naiveisocurvature}
\end{equation}
where we used $\Delta N_* = 55$ and $N_{\rm{tot}} = 76.5$ $e$-folds. This result agrees extremely well with the fully numerical solution Eq.~(\ref{eq:isocurvaturecons1}). However, for the Starobinsky and T-models of inflation, considered in this work, the Hubble rate varies with time. One can therefore evaluate Eq.~(\ref{eq:naiveisocurvature}) in terms of the Hubble scale at the end of inflation,
\begin{align}
    \label{eq:limpurgrav}
    &m_{\chi} \gtrsim 1.18 H_{\rm{end}} \qquad  (\text{Starobinsky})\, , ~\notag \\
    &m_{\chi} \gtrsim 1.41 H_{\rm{end}} \qquad (\text{T-model}) \, .
\end{align}
These results agree extremely well with the fully numerical result~(\ref{eq:numconstraints1}). Assuming that the bare DM mass is negligible compared to the contribution induced by the inflaton $m_{\rm{eff}}^2  \simeq \sigma \phi^2$, the bound~(\ref{eq:naiveisocurvature}) corresponds to
\begin{align} \label{eq:naiveisocurvaturesigmaoverlambda}
   \dfrac{\sigma}{\lambda} \, > \, 0.006 \, (0.02)~~~ {\textrm{for Starobinsky~(T-model)}}\, , ~\nonumber\\
\end{align}
which is in good agreement with the fully numerical result~(\ref{eq:numconstraints2}).
\addcontentsline{toc}{section}{References}
\bibliographystyle{utphys}
\bibliography{references}

\end{document}